\documentclass[10pt,a4paper,fleqn,twocolumn]{article}
\usepackage[utf8]{inputenc}
\usepackage[english]{babel}
\usepackage{amsmath}
\usepackage{amsfonts}
\usepackage{amssymb}
\usepackage{graphicx}
\usepackage{fourier}
\usepackage{bm}
\usepackage[left=1cm,right=1cm,top=2cm,bottom=2cm]{geometry}
\usepackage{enumitem}

\usepackage{hyperref} % \usepackage[hidelinks]{hyperref}
\hypersetup{
    colorlinks=true,
    linkcolor=blue,   
    filecolor=blue,  
    urlcolor=blue,
    citecolor=blue,
    breaklinks=true
    }
\usepackage[authoryear,round]{natbib}
\usepackage{doi}
\usepackage{orcidlink}

\title{Effective strain energy of soft laminates:\\ Beyond neo-Hookean elasticity}
\author{Harold Berjamin \textsuperscript{a,b,*} \orcidlink{0000-0001-6637-8485} and Stephan Rudykh \textsuperscript{a} \orcidlink{0000-0002-4568-8326} \\ ~ \\
\begin{minipage}{\textwidth}
\centering\footnotesize
\textsuperscript{a}School of Mathematical and Statistical Sciences, University of Galway, Galway, H91 TK33, Ireland \\ 
\textsuperscript{b}Mathematics Applications Consortium for Science and Industry (MACSI),\\ Department of Mathematics and Statistics, University of Limerick, Limerick, V94 T9PX, Ireland \\
\textsuperscript{*}Corresponding author. Present address: School of Engineering, University of Galway, Galway, H91 TK33, Ireland. \\ Email address: \href{mailto:harold.berjamin@universityofgalway.ie}{harold.berjamin@universityofgalway.ie}
\end{minipage}}
\date{}

\begin{document}

\twocolumn[
  \begin{@twocolumnfalse}

\maketitle

\begin{abstract}
\noindent
Composite laminates can be designed to reach desired engineering properties such as low weight and high stiffness. This study concerns periodic media made of alternating layers of soft rubber-like solids. Assuming incompressible and perfectly bonded elastic layers, the effective mechanical response of the laminate is estimated by means of homogenisation theory. This involves computing the spatial average of the energy stored in the layers, each of which is in equilibrium and homogeneously deformed. The study is conducted for layers whose elastic energy depends on two scalar invariants, in general. Beyond neo-Hookean solids, explicit analytical expressions are obtained for a special class of Mooney--Rivlin materials, whereas approximate expressions are obtained for Mooney--Rivlin laminates and for Yeoh laminates. These stratified media can be represented by a transversely isotropic elasticity theory at the macroscopic scale. Based on these results, an approximate effective energy for laminates whose layers combine Gent-type and Mooney--Rivlin behaviour is postulated. It is then used to estimate the macroscopic instability threshold under pure shear numerically. Results show that the specific mechanical response of individual layers has negligible influence on the macroscopic stability limit if a certain proportionality relationship between their energy functions holds. Furthermore, in a pre-stressed layered medium under pure shear, the influence of the individual mechanical response of the layers on the shear wave velocity is found to be comparable to that observed in homogeneous media, within the same range of investigated situations. Finally, more significant effects may be observed in non-proportional laminates.
\end{abstract}

\end{@twocolumnfalse}
]

%%%%%%%%%%%%%%%%%%%%%%%%%%%%%%%%%%%%%%%%%%%%%%%%%%%%%%%
\section{Introduction}\label{sec:Intro}
%%%%%%%%%%%%%%%%%%%%%%%%%%%%%%%%%%%%%%%%%%%%%%%%%%%%%%%

Composite laminates are commonly found in Engineering applications, for instance in construction and in manufacturing \citep{gibson16}. The simplest composite structure is the \emph{bi-laminate} consisting of two alternating layers of material. One example is plywood, which is composed of thin layers of wood that are glued together, usually so that the grain of each layer is perpendicular to that of the adjacent layers. Thanks to improvements in this technique that dates back to ancient Egypt, modern plywood can achieve high flexural strength for a low mass density \citep{ansell15}.

The present paper is dedicated to the study of soft bi-laminates whose layers are perfectly bonded. The layers are made of rubber-like materials which are able to sustain large elastic deformations. Although bi-layered soft composites are conceptually simple, their mechanical response is already very rich. For instance, \citet{yue94} studied the wrinkling of neo-Hookean plates under plane strain, showing that the critical wrinkling load of a bi-layered plate is highly dependent on the stiffness contrast between the layers. Similarly, \citet{roccabianca10} found that the mechanical behaviour of a neo-Hookean medium subjected to bending is strongly affected by the presence of multiple layers with a distinct stiffness. Along the same lines, the dynamic response of a layered beam was studied by \citet{parnell07}, and a study of the anticlastic bending of a soft laminated beam was carried out by \citet{bacciocchi22}.

For a soft composite made of many layers, it may be useful in practice to resort to accurate macroscopic theories that describe the mechanical response of the material without having to consider each layer individually. In this regard, the effective mechanical response of a periodic bi-layered composite was analysed by \citet{debotton05} using homogenisation theory. This way, by means of spatial averaging, the effective strain energy function of a neo-Hookean bi-laminate can be determined analytically \citep[Appendix A]{agoras09}. The resulting macroscopic theory is a transversely isotropic extension of neo-Hookean elasticity accounting for the orientation of the layers.

This classical result obtained for incompressible neo-Hookean laminates was used in a variety of applications, ranging from studies on material stability to studies on wave propagation \citep{galich17}, as well as studies on the electro-active or magneto-active response of soft laminates \citep{bertoldi11,rudykh11}. Related results were obtained for compressible elastic laminates by \citet{lopezpamies09}, who relied on the tangent second-order homogenisation method to study the evolution of the microstructure. The formation of microstructural patterns in neo-Hookean laminates is investigated in the works by \citet{furer18} and by \citet{li22}.

Most of the analytical homogenisation results available in the literature were obtained for neo-Hookean materials, whose strain energy is a linear function of the first strain invariant only ($I_1$). Based on the classical homogenisation procedure mentioned above, \citet{berjamin25} derive a polynomial expression of the effective energy \eqref{WYeoh} for soft laminates made of incompressible Yeoh materials with two terms in the expansion of the energy (i.e., whose energy is described by a quadratic function of the first strain invariant). In the present study, we use the same approach to generalise this result to a wider class of constitutive laws, namely incompressible materials whose elastic energy depends on both strain invariants ($I_1$ and $I_2$), as well as Yeoh materials with three terms in the expansion of the strain energy (i.e., whose energy is described by a cubic polynomial function of the first strain invariant). The importance of accounting for the $I_2$-dependence of the energy is highlighted in several works, see for instance \citet{kuhl24}.

\begin{table*}
	\caption{ Effective elastic energy of soft elastic laminates. A layer's shear modulus is $\mathcal{G}^{(\alpha)}$, while the invariants $I_k^{(\alpha)}$ are defined in Eqs.~\eqref{InvarMicro} and \eqref{InvarHom}. The first line is an exact result obtained for special Mooney--Rivlin laminates such that $\delta^{(\alpha)}=\delta$, generalising the neo-Hookean result by \citet{debotton05} to $\delta \neq 0$. The second line is an asymptotic result obtained for Mooney--Rivlin laminates for which $\delta^{(1)}\simeq \delta^{(2)} \ll 1$. The last line is an asymptotic result obtained for Yeoh laminates where the parameters of nonlinearity $b^{(\alpha)}$ and $d^{(\alpha)}$ are small. This expression generalises the one by \citet{berjamin25} to three-term Yeoh laminates with $d^{(\alpha)} \neq 0$. The linear average of the shear moduli is $\langle\mathcal{G}\rangle$, their harmonic average is $\breve{\mathcal G}$, and other averaged quantities are calculated according to Eq.~\eqref{AverNotation}. The quantities $D$, $E$, $K$, $L$ are defined in Eqs.~\eqref{IdDE}-\eqref{IdKL}. \label{tab:Expr} }
	\begin{minipage}{\textwidth}
 	\centering
	\renewcommand\footnoterule{{\hrule height 0pt}}
	{\renewcommand{\arraystretch}{1.2}
	\renewcommand{\thempfootnote}{\fnsymbol{mpfootnote}}
	\rmfamily
	\begin{tabular*}{\linewidth}{ lll }
	\hline
	Strain energy $W^{(\alpha)}$ of the layers ($\alpha=1,2$) & Effective strain energy $\widetilde{W}$ of the bi-laminate  & \begin{tabular}{c} Eq. \\ E/A\footnote{E: exact result; A: asymptotic result.} \end{tabular} \\
	\hline 
	$\frac12 (1-\delta) \mathcal{G}^{(\alpha)} ( I_1^{(\alpha)}-3 ) + \frac12 \delta \mathcal{G}^{(\alpha)} (I_2^{(\alpha)} - 3)$ & $\frac12 (1-\delta) \langle\mathcal{G}\rangle D + \frac12 \delta \langle\mathcal{G}\rangle E + \frac12 (1-\delta) \breve{\mathcal G} K + \frac12 \delta \breve{\mathcal G} L$ & \begin{tabular}{c} \eqref{WMRSimplEff} \\ E \end{tabular} \\
	$\frac12 (1 - \delta^{(\alpha)}) \mathcal{G}^{(\alpha)} (I_1^{(\alpha)}-3) + \frac12 \delta^{(\alpha)} \mathcal{G}^{(\alpha)} (I_2^{(\alpha)}-3)$ & $\frac12 \langle \mathcal{G} (1-\delta)\rangle D + \frac12 \langle \mathcal{G} \delta\rangle E + \frac12 \breve{\mathcal G}^2 \langle \mathcal{G}^{-1} (1-\delta) \rangle K + \frac12 \breve{\mathcal G}^2 \langle \mathcal{G}^{-1} \delta \rangle L$ & \begin{tabular}{c} \eqref{WMREff} \\ A \end{tabular} \\
	${\begin{aligned} &\tfrac12 \mathcal{G}^{(\alpha)} (I_1^{(\alpha)} - 3) \\
		&\; + \tfrac14 b^{(\alpha)} \mathcal{G}^{(\alpha)} (I_1^{(\alpha)} - 3)^2 \\
		&\; + \tfrac1{12} d^{(\alpha)} \mathcal{G}^{(\alpha)} (I_1^{(\alpha)} - 3)^3
	\end{aligned}}$ & ${\begin{aligned}
		&\tfrac12 \langle\mathcal G\rangle D + \tfrac12 \breve{\mathcal G} K \\
		&\; + \tfrac14 \langle\mathcal{G} b\rangle D^2 + \tfrac12 \breve{\mathcal{G}}^2 \langle\mathcal{G}^{-1} b\rangle DK + \tfrac14 \breve{\mathcal{G}}^4 \langle\mathcal{G}^{-3} b\rangle K^2 \\
	&\; + \tfrac1{12} \langle\mathcal{G}d\rangle D^3 + \tfrac1{4} \breve{\mathcal G}^2 \langle\mathcal{G}^{-1}d\rangle D^2K + \tfrac1{4} \breve{\mathcal G}^4 \langle\mathcal{G}^{-3}d\rangle DK^2 + \tfrac1{12} \breve{\mathcal G}^6 \langle\mathcal{G}^{-5}d\rangle K^3
	\end{aligned}}$ & \begin{tabular}{c} ~ \\ \eqref{WGenNH} \\ A \\ ~ \end{tabular} \\
	\hline
	\end{tabular*}}
	\end{minipage}
\end{table*}

Using analytical expansions, we provide an asymptotic expression for the effective energy of Mooney--Rivlin laminates, see Eqs.~\eqref{WMRSimplEff} and \eqref{WMREff}. Furthermore, we provide an asymptotic expression of the effective energy \eqref{WGenNH} for Yeoh laminates, whose layers are described by a general cubic energy function of the first strain invariant (i.e., a polynomial function with linear, quadratic and cubic terms). These effective theories summarised in Table~\ref{tab:Expr} are then used to study the influence of the constitutive response of individual layers on the macroscopic instability threshold and on the speed of shear waves under pure shear. If a certain proportionality relationship between the energy functions describing both phases holds, results show that the specific mechanical response of the layers has negligible influence on the value of the macroscopic instability threshold, within the range of investigated cases. Furthermore, under this proportionality assumption, the study shows that the behaviour of the constituent layers affects the speed of incremental shear waves to an extent comparable to that observed in homogeneous media. Further results suggest that these observations may not hold outside the proportionality hypothesis.

The paper is organised as follows. Section~\ref{sec:Problem} describes the layers and their elastic response, as well as the homogenisation procedure. Section~\ref{sec:Weff} presents the analytical results concerning the effective strain energy of soft laminates, whose accuracy is assessed in Section~\ref{sec:Valid}. Based on these results, Section~\ref{sec:Insta} is dedicated to the study of incremental motions under pure shear. Concluding remarks are provided in Section~\ref{sec:Conclu}. The Appendix~\ref{app:Acoust} provides detailed expressions involving the components of the acoustic tensor.

%%%%%%%%%%%%%%%%%%%%%%%%%%%%%%%%%%%%%%%%%%%%%%%%%%%%%%%
\section{Problem statement}\label{sec:Problem}
%%%%%%%%%%%%%%%%%%%%%%%%%%%%%%%%%%%%%%%%%%%%%%%%%%%%%%%
\subsection{Microscopic description of soft elastic laminates}
%%%%%%%%%%%%%%%%%%%%%%%%%%%%%%%%%%%%%%%%%%%%%%%%%%%%%%%

\begin{figure}
    \centering
	\includegraphics{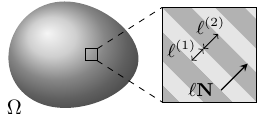}
	\caption{Undeformed layered medium and view of its microstructure. \label{fig:Laminate}}
\end{figure}

Consider a biphasic composite material made of perfectly bonded alternating layers, with a given initial thickness $\ell^{(\alpha)}$, where $\alpha = \lbrace 1, 2\rbrace$ indicates each phase. This way, the microstructure has initial spatial period $\ell = \ell^{(1)} + \ell^{(2)}$ along the lamination direction orientated by the unit vector ${\bf N}$ (such that ${\bf N} \cdot {\bf N} = |{\bf N}|^2 = 1$), as depicted in Figure~\ref{fig:Laminate}. The layers are made of rubber-like elastic solids, which are assumed isotropic and incompressible. In the present study, we assume that the layer thickness is much smaller than the characteristic spatial scale of the deformation, which is itself smaller than the sample size.

Material points are described by their initial location $\bf X$ in the undeformed medium (delimited by the boundaries of a given material domain $\Omega$), to which corresponds the vector $\bf x$ indicating their current position in the deformed medium. The deformation gradient tensor is defined by ${\bf F} = \partial {\bf x}/\partial {\bf X}$, which is the gradient of the map ${\bf X} \mapsto {\bf x}$. \emph{Incompressible} media do not allow for volume change, so that the constraint $\det {\bf F} \equiv 1$ is enforced everywhere in the material.

The constitutive behaviour of the phase $\alpha$ is characterised by the strain energy $W^{(\alpha)}({\bf F})$, which is a given function of $\bf F$ involving relevant material parameters. Introducing the characteristic function $\chi^{(\alpha)}$ of the regions occupied by the phase $\alpha$ (such that $\chi^{(\alpha)}({\bf X}) = 1$ if $\bf X$ is located in the phase $\alpha$, and $\chi^{(\alpha)}({\bf X}) = 0$ otherwise), the strain energy function can be expressed as
\begin{equation}
	W = \chi^{(1)} W^{(1)} + \chi^{(2)} W^{(2)} ,
	\label{WMicro}
\end{equation}
which accounts for the local variations of the material properties. Due to the periodicity of the microstructure, the characteristic functions $\chi^{(\alpha)}$ are $\ell$-periodic along the material direction $\bf N$, i.e., $\chi^{(\alpha)}({\bf X}+\ell{\bf N}) = \chi^{(\alpha)}({\bf X})$ for all $\bf X$, and we have the relationship $\chi^{(2)} = 1-\chi^{(1)}$.

In incompressible solids, the first Piola--Kirchhoff stress tensor is then defined by
\begin{equation}
	{\bf P} = \frac{\partial W}{\partial \bf F} - p {\bf F}^{-\text{T}} ,
	\label{StressMicroGen}
\end{equation}
where $p({\bf X})$ is a Lagrange multiplier introduced to enforce the constraint of incompressibility at every point, and \textsuperscript{T} indicates the transpose. The Cauchy stress tensor $\bf T$ can then be evaluated using the definition ${\bf T} = {\bf P} {\bf F}^{\text{T}}$, see the monograph by \citet{holzapfel00}. The balance of angular momentum requires that ${\bf T}$ is a symmetric tensor, or equivalently, that ${\bf P} {\bf F}^{\text{T}} = {\bf F} {\bf P}^{\text{T}}$.

In isotropic elastic solids, the strain energy is function of at most three independent scalar quantities which are related to a given measure of deformation. Due to incompressibility, the number of such independent scalar quantities is reduced to two. Here, the strain energy $W^{(\alpha)}$ in each layer is assumed to be a general analytical function of the principal strain invariants
\begin{equation}
	I_1 = \text{tr}\, {\bf C} , \quad I_2 = \text{tr}\, {\bf C}^{-1} ,
	\label{InvarMicro}
\end{equation}
where ${\bf C} = {\bf F}^{\text{T}} {\bf F}$ is the right Cauchy--Green strain tensor. Using the above definitions, we find
\begin{equation}
	{\bf P} = \left( 2W_1 {\bf B} - 2W_2 {\bf B}^{-1} - p {\bf I} \right) {\bf F}^{-\text{T}} ,
	\label{StressMicro}
\end{equation}
where ${\bf B} = {\bf F} {\bf F}^{\text{T}}$ is the left Cauchy--Green strain tensor. The functions $W_k = \partial W / \partial I_k$ are partial derivatives of the strain energy $W$, and $\bf I$ is the second-order identity tensor. Due to incompressibility, both strain tensors ${\bf B}$ and ${\bf C}$ have unit determinant.

The present formulation is valid for laminates whose layers are isotropic hyperelastic solids, which are only subjected to the constraint of incompressibility (i.e., absence of volume change). Therefore, it encompasses many common models of the literature which are used to describe the elastic deformations of soft rubber-like solids. Typical examples of such material models are the incompressible neo-Hookean, Mooney--Rivlin, Yeoh, Fung--Demiray and Gent theories, among other similar models. For a complete overview of rubber and polymer elasticity, one may refer the interested reader to reviews of the topic \citep{marckmann06,steinmann12,destrade17,dal21,he22,ricker23}. For recent developments, see for instance the incompressible models by \citet{bhat25} and by \citet{shah26}, as well as the compressible models by \citet{pelliciari23} and by \citet{anssari25}.

%%%%%%%%%%%%%%%%%%%%%%%%%%%%%%%%%%%%%%%%%%%%%%%%%%%%%%%
\subsection{Macroscopic response of soft elastic laminates}
%%%%%%%%%%%%%%%%%%%%%%%%%%%%%%%%%%%%%%%%%%%%%%%%%%%%%%%

For soft laminates with isotropic phases of neo-Hookean type or of Yeoh type, it was shown in earlier works that the macroscopic elastic response of the laminate corresponds to that of a specific transversely isotropic and incompressible medium \citep{berjamin25}. In what follows, we generalise these results to a wider class of constitutive models, and we provide explicit analytical estimates of the effective mechanical response of the laminate.

The macroscopic elastic response of an incompressible soft solid with one special material direction (such as the lamination direction, ${\bf N}$) can be described within the framework of transversely isotropic elasticity. To do so, we introduce the macroscopic deformation gradient, $\overline{\bf F}$, which is defined as the volume average $\langle {\bf F} \rangle$ of the deformation gradient over the whole material domain $\Omega$ \citep{hill72},
\begin{equation}
	\overline{\bf F} = \langle {\bf F} \rangle = \frac{\int_\Omega {\bf F}\, \text{d}V}{\int_\Omega \text{d}V} .
	\label{FMacro}
\end{equation}
For the description of transversely isotropic solids, we introduce the four scalar invariants
\begin{equation}
	\tilde I_1 = \text{tr}\, \overline{\bf C} , \quad \tilde I_2 = \text{tr}\, \overline{\bf C}^{-1} , \quad \tilde I_4 = {\bf N}\cdot \overline{\bf C} {\bf N} , \quad \tilde I_5 = {\bf N}\cdot \overline{\bf C}^{-1} {\bf N} ,
	\label{InvarMacro}
\end{equation}
which are the arguments of the macroscopic strain energy function, $\widetilde{W}$, and the tensor $\overline{\bf C} = \overline{\bf F}^\text{T} \overline{\bf F}$ represents the macroscopic deformation.

The volume average of Piola--Kirchhoff stress $\overline{\bf P} = \langle {\bf P} \rangle$ can be evaluated as $\overline{\bf P} = \partial \widetilde{W} / \partial \overline{\bf F}$, cf. \citet{ogden74}, or $\overline{\rm P}_{ia} = \partial \widetilde{W} / \partial \overline{\rm F}_{ia}$, componentwise. Thus, using the formulas in the Appendix~\ref{app:Acoust}, we find
\begin{equation}
	\begin{aligned}
	\overline{\bf P} &= \left(2 \widetilde{W}_1 \overline{\bf B} - 2 \widetilde{W}_2 \overline{\bf B}^{-1} % \right.\\
%&\quad\left.\phantom{0} 
+ 2 \tilde I_4 \widetilde{W}_4 {\bf n} \otimes {\bf n} - 2 \tilde I_5 \widetilde{W}_5 {\bf m} \otimes {\bf m} - \overline{p}\, {\bf I} \right) \overline{\bf F}^{-\text{T}} , 
	\end{aligned}
	\label{StressMacro}
\end{equation}
where $\overline{\bf B} = \overline{\bf F}\overline{\bf F}^\text{T}$. The vectors $\bf n$ and $\bf m$ are unit vectors such that
\begin{equation}
	{\bf n} = \frac{\overline{\bf F} {\bf N}}{| \overline{\bf F} {\bf N} |} = \frac{\overline{\bf F} {\bf N}}{\sqrt{\tilde I_4}} , \quad
	{\bf m} = \frac{\overline{\bf F}^{-\text{T}} {\bf N}}{| \overline{\bf F}^{-\text{T}} {\bf N} |} = \frac{\overline{\bf F}^{-\text{T}} {\bf N}}{\sqrt{\tilde I_5}}  .
	\label{VecN}
\end{equation}
Here, the scalar field $\overline p$ is a Lagrange multiplier to enforce the incompressibility constraint at the macroscopic scale, $\otimes$ is the tensor product (dyadic, or outer product), and similar notation is used as above. In particular, the functions $\widetilde{W}_k$ correspond to the partial derivatives $\partial \widetilde{W}/\partial \tilde I_k$. The expression of the stress \eqref{StressMicro} in the isotropic case is recovered for $\widetilde{W}_4 = \widetilde{W}_5 = 0$.

On the one hand, the effective direction of anisotropy in the deformed medium is indicated by the unit vector $\bf m$, consistently with Nanson's formula \citep{holzapfel00}. On the other hand, the unit vector $\bf n$ is indicating the current direction of a material point whose initial position was described by a vector along $\bf N$ (or equivalently, the direction of a curve whose material tangent vector is $\bf N$), and we have the property ${\bf n} \cdot {\bf m} = (\tilde I_4 \tilde I_5)^{-1/2}$. Therefore, the unit vectors $\bf n$ and $\bf m$ are only codirectional if $\tilde I_4 = \tilde I_5^{-1}$, which is not expected to be true in general.

The set of invariants \eqref{InvarMacro} can be connected to other common choices across the literature by making use of the Cayley--Hamilton identity for the right Cauchy--Green tensor $\overline{\bf C}$,
\begin{equation}
	\overline{\bf C}^{-1} = \overline{\bf C}^2 - \tilde I_1 \overline{\bf C} + \tilde I_2 {\bf I}, 
	\label{CayleyHamilton}
\end{equation}
which implies that
\begin{equation}
	\tilde I_2 = \tfrac12 \left( \tilde I_1^2 - \text{tr}(\overline{\bf C}^2) \right) , \quad \tilde I_5 = {\bf N}\cdot \overline{\bf C}^2 {\bf N} - \tilde I_1 \tilde I_4 + \tilde I_2 .
	\label{InvarRel}
\end{equation}
In some related works, the second invariant is defined by Eq.~\eqref{InvarRel}\textsubscript{1} and the fifth invariant is defined as ${\bf N}\cdot \overline{\bf C}^2 {\bf N}$ instead of the present definition \eqref{InvarMacro}, see for instance \citet{holzapfel00} and \citet{destrade12}. Due to the relationships \eqref{InvarRel}, these choices are equivalent, in general \citep{merodio02}. For other sets of invariants motivated by physical considerations, the interested reader is referred to works by \citet{criscione01} as well as \citet{shariff08}.

%%%%%%%%%%%%%%%%%%%%%%%%%%%%%%%%%%%%%%%%%%%%%%%%%%%%%%%
\subsection{Homogenisation procedure}
%%%%%%%%%%%%%%%%%%%%%%%%%%%%%%%%%%%%%%%%%%%%%%%%%%%%%%%

It remains now to connect the effective response of the laminate described by the strain energy function $\widetilde{W}(\tilde I_1, \dots , \tilde I_5)$ to the microscopic properties of the layers, which are described by the strain energy \eqref{WMicro}. To do so, the following minimisation problem is formulated \citep{hill72},
\begin{equation}
	\widetilde{W} = \min_{{\bf F}\in \Sigma} \, \langle W \rangle ,
	\label{WMini}
\end{equation}
where $\langle \cdot \rangle$ denotes the volume averaging operator, see Eq.~\eqref{FMacro}. In this minimisation problem, the deformation gradient tensor ${\bf F}$ spans the set $\Sigma$ of admissible motions which are compatible with the macroscopic deformation gradient $\overline{\bf F}$ on the boundaries of $\Omega$, where the homogeneous motion ${\bf x} = \overline{\bf F}{\bf X}$ is imposed.

A solution of the Euler--Lagrange equations for the minimisation problem \eqref{WMini} can be computed by assuming homogeneous deformations in the layers \citep{debotton05}, that is ${\bf F} = \chi^{(1)} {\bf F}^{(1)} + \chi^{(2)} {\bf F}^{(2)}$, where the phase-related deformation gradients ${\bf F}^{(\alpha)}$ are independent of $\bf X$. Due to incompressibility, the latter must satisfy the constraint $\det {\bf F}^{(\alpha)} \equiv 1$. Then, the relevant jump conditions for the displacement field and the normal stresses across adjacent layers are enforced. In principle, the energy function so-obtained is valid up to the onset of macroscopic instability, see also \citet{lopezpamies09} where a more sophisticated homogenisation approach is used.

The continuity of the displacement field at a material interface between two adjacent layers requires that ${\bf F}^{(1)} {\bf X} = {\bf F}^{(2)}{\bf X}$ for every material point ${\bf X}$ belonging to that interface. This condition may be rewritten as the requirement ${\bf F}^{(1)} {\bm \phi} = {\bf F}^{(2)}{\bm \phi}$ holding for every vector ${\bm \phi}$ within the plane orthogonal to $\bf N$, which tells us that the kernel of ${\bf F}^{(2)}-{\bf F}^{(1)}$ is of dimension two, at least. Thus, we have the representation formula \citep{debotton05}
\begin{equation}
	{\bf F}^{(2)}-{\bf F}^{(1)} = {\bm\theta} \otimes {\bf N}, \quad {\bm\theta} = \left({\bf F}^{(2)}-{\bf F}^{(1)}\right) {\bf N} .
	\label{RepDef}
\end{equation}
The macroscopic deformation gradient \eqref{FMacro} becomes $\overline{\bf F} = \nu^{(1)}{\bf F}^{(1)} + \nu^{(2)}{\bf F}^{(2)}$, where $\nu^{(\alpha)} = \langle \chi^{(\alpha)} \rangle = \ell^{(\alpha)}/\ell$ is the initial volume fraction of phase $\alpha$, and $\nu^{(1)}+\nu^{(2)}=1$.

From the above observations, it follows that
\begin{equation}
	{\bf F}^{(\alpha)} = \overline{\bf F} + \tau^{(\alpha)} {\bm\theta} \otimes {\bf N} , \quad
	\tau^{(\alpha)} = \begin{cases}
	-\nu^{(2)}, \;\text{if}\; \alpha = 1,\\
	\phantom{-}\nu^{(1)}, \;\text{if}\; \alpha = 2.
	\end{cases}
	\label{FdeBotton}
\end{equation}
To enforce incompressibility in each phase, we require
\begin{equation}
	\det {\bf F}^{(\alpha)} = (\det \overline{\bf F}) \left(1 + \tau^{(\alpha)} \overline{\bf F}^{-1}{\bm\theta} \cdot {\bf N}\right) = 1 ,
	\label{DetdeBotton}
\end{equation}
from where we deduce that the scalar product $\overline{\bf F}^{-1}{\bm\theta} \cdot {\bf N} = {\bm\theta} \cdot \overline{\bf F}^{-\text{T}} {\bf N}$ vanishes, and in particular, that ${\bm\theta} \cdot {\bf m} = 0$, see Eq.~\eqref{VecN}. The macroscopic strain energy \eqref{WMini} now takes the form
\begin{equation}
	\widetilde{W} = \min_{{\bm \theta} \bot {\bf m}} \left\lbrace \nu^{(1)} W^{(1)} + \nu^{(2)} W^{(2)} \right\rbrace .
	\label{WMiniDeBotton}
\end{equation}

The phase-related energy contributions $W^{(\alpha)}$ are functions of the two invariants defined in Eq.~\eqref{InvarMicro}, which are deduced from the expression \eqref{FdeBotton} of ${\bf F}^{(\alpha)}$ and from the identity \citep{lopezpamies09}
\begin{equation}
	({\bf F}^{(\alpha)})^{-\text{T}} = \overline{\bf F}^{-\text{T}} - \tau^{(\alpha)} \overline{\bf F}^{-\text{T}}{\bf N} \otimes \overline{\bf F}^{-1} {\bm\theta} .
	\label{FdeBottonInv}
\end{equation}
The above quantity corresponds to the inverse transpose of the tensor ${\bf F}^{(\alpha)}$ which is near the invertible tensor $\overline{\bf F}$, in that the difference ${\bf F}^{(\alpha)} - \overline{\bf F}$ is of rank one, see Eq.~\eqref{FdeBotton}.
Using these formulas in the evaluation of the strain tensor ${\bf B}^{(\alpha)} = {\bf F}^{(\alpha)}{\bf F}^{(\alpha)\text{T}}$ and of its inverse, we find that
\begin{equation}
	\begin{aligned}
		& {\bf B}^{(\alpha)} = \overline{\bf B} + \tau^{(\alpha)} (\overline{\bf F}{\bf N} \otimes {\bm\theta} + {\bm\theta}\otimes \overline{\bf F}{\bf N}) + (\tau^{(\alpha)})^2 {\bm\theta} \otimes {\bm\theta} , \\
		& ({\bf B}^{(\alpha)})^{-1} = \overline{\bf B}^{-1} - \tau^{(\alpha)} (\overline{\bf F}^{-\text{T}}{\bf N} \otimes \overline{\bf B}^{-1} {\bm\theta} + \overline{\bf B}^{-1} {\bm\theta}\otimes \overline{\bf F}^{-\text{T}}{\bf N}) \\
&\qquad\qquad + (\tau^{(\alpha)})^2 |\overline{\bf F}^{-1} {\bm\theta}|^2 \overline{\bf F}^{-\text{T}}{\bf N} \otimes \overline{\bf F}^{-\text{T}}{\bf N} .
	\end{aligned}
	\label{StrainPhase}
\end{equation}
It follows that the principal strain invariants in each phase satisfy
\begin{equation}
	\begin{aligned}
		& I_1^{(\alpha)} = \tilde I_1 + 2\tau^{(\alpha)} \overline{\bf F} {\bf N}\cdot {\bm\theta} + (\tau^{(\alpha)})^2 |{\bm\theta}|^2 , \\
		& I_2^{(\alpha)} = \tilde I_2 - 2\tau^{(\alpha)} \overline{\bf F}^{-\text{T}} {\bf N} \cdot \overline{\bf B}^{-1}{\bm\theta}  + (\tau^{(\alpha)})^2 \tilde I_5 |\overline{\bf F}^{-1}{\bm\theta}|^2 ,
	\end{aligned}
	\label{InvarHom}
\end{equation}
which are the traces of the tensors \eqref{StrainPhase}.

The Euler--Lagrange equation associated with the minimisation problem \eqref{WMiniDeBotton} is the traction continuity equation ${\bf t}^{(1)} = {\bf t}^{(2)}$, where the tractions ${\bf t}^{(\alpha)} = {\bf P}^{(\alpha)}{\bf N}$ are deduced from the first Piola--Kirchhoff stresses ${\bf P}^{(\alpha)} = \partial W^{(\alpha)} / \partial {\bf F}^{(\alpha)} - p^{(\alpha)} ({\bf F}^{(\alpha)})^{-\text{T}}$ in each phase \citep{debotton05}. Noting that $({\bf F}^{(\alpha)})^{-\text{T}} {\bf N} = \overline{\bf F}^{-\text{T}} {\bf N}$ due to the relationships \eqref{DetdeBotton} and \eqref{FdeBottonInv}, the continuity requirement for the tractions is then used to determine the optimal vector ${\bm\theta}$. According to Eq.~\eqref{StressMicro} and to the above identities, we have
\begin{equation}
	\begin{aligned}
	&{\bf t}^{(\alpha)} = {\bf s}^{(\alpha)} - p^{(\alpha)} \overline{\bf F}^{-\text{T}} {\bf N}, \\
	&{\bf s}^{(\alpha)} = \left( 2 W_1^{(\alpha)} {\bf B}^{(\alpha)}  - 2 W_2^{(\alpha)}  ({\bf B}^{(\alpha)})^{-1} \right) \overline{\bf F}^{-\text{T}} {\bf N} .
	\end{aligned}
	\label{Traction}
\end{equation}
Here, the coefficients $W_k^{(\alpha)} = \partial W^{(\alpha)} / \partial I_k^{(\alpha)}$ are partial derivatives of the strain energy, and the coefficients $p^{(\alpha)}$ are the Lagrange multipliers to enforce incompressibility in each layer.

The next step of this procedure is to determine the pressure jump, $p^{(2)} - p^{(1)}$. To do so, we evaluate the scalar product of Eq.~\eqref{Traction} with $\overline{\bf F}^{-\text{T}} {\bf N}$,
\begin{equation}
	{\bf N}\cdot \overline{\bf F}^{-1} {\bf t}^{(\alpha)} = {\bf N}\cdot \overline{\bf F}^{-1} {\bf s}^{(\alpha)} - p^{(\alpha)} \tilde I_5 .
	\label{Press}
\end{equation}
Enforcing the continuity of tractions ${\bf t}^{(1)} = {\bf t}^{(2)}$ in Eq.~\eqref{Press} allows to determine the expression of the pressure jump,
\begin{equation}
	p^{(2)} - p^{(1)} = \tilde I_5^{-1} {\bf N}\cdot \overline{\bf F}^{-1} \left( {\bf s}^{(2)}-{\bf s}^{(1)} \right) .
\end{equation}
Then, this expression is inserted in the traction continuity equation ${\bf t}^{(1)} = {\bf t}^{(2)}$ deduced from Eq.~\eqref{Traction} to eliminate the Lagrange multipliers $p^{(\alpha)}$. Doing so, the continuity of tractions at the layer interfaces reduces to the condition
\begin{equation}
	\widehat{\bf I} \left({\bf s}^{(2)} - {\bf s}^{(1)}\right) = {\bf 0} ,
	\quad
	\widehat{\bf I} = {\bf I} - {\bf m} \otimes {\bf m},
	\label{TractionFinal}
\end{equation}
where $\widehat{\bf I}$ is the projection tensor onto the plane orthogonal to the unit vector ${\bf m} = \tilde I_5^{-1/2} \overline{\bf F}^{-\text{T}} {\bf N}$ introduced in Eq.~\eqref{VecN}.

The expression of the projected tractions $\widehat{\bf I} \, {\bf s}^{(\alpha)}$ deduced from Eq.~\eqref{Traction} can be obtained by making use of Eq.~\eqref{StrainPhase} and of the orthogonality property ${\bm\theta} \cdot {\bf m} = 0$. Doing so, we find
\begin{equation}
	\widehat{\bf I} \, {\bf s}^{(\alpha)} = {\bf A}^{(\alpha)} \left( {\bf u} + \tau^{(\alpha)} {\bm \theta} \right) ,  
	\label{TractionProj}
\end{equation}
where
\begin{equation}
	{\bf A}^{(\alpha)} = 2 W_1^{(\alpha)} {\bf I} + 2 W_2^{(\alpha)} \tilde I_5 \widehat{\bf I} \, \overline{\bf B}^{-1} , \quad
	{\bf u} = \widehat{\bf I}\, \overline{\bf F} {\bf N} ,
	\label{VecU}
\end{equation}
which follows from the relationship $\tilde I_5 \widehat{\bf I} \, \overline{\bf B}^{-1} {\bf u} = -\widehat{\bf I} \, \overline{\bf B}^{-1} \overline{\bf F}^{-\text{T}} {\bf N}$. It is worth noting that the matrices ${\bf A}^{(\alpha)}$ depend on $\bm\theta$ through the response functions $W_1^{(\alpha)}$ and $W_2^{(\alpha)}$, see the definition of the invariants \eqref{InvarHom}. Hence, solving the traction continuity equation \eqref{TractionFinal} seems very challenging, in general, and approximate methods of resolution are needed (e.g., numerical methods). Once $\bm\theta$ has been determined, the invariants \eqref{InvarHom} and the effective strain energy \eqref{WMiniDeBotton} can be evaluated, in principle{\,---\,}again, these calculations are not always straightforward. Nevertheless, exact or approximate analytical expressions of the effective energy can be obtained in some special cases, as described hereinafter.

%%%%%%%%%%%%%%%%%%%%%%%%%%%%%%%%%%%%%%%%%%%%%%%%%%%%%%%
\section{Homogenisation results for various constitutive laws}\label{sec:Weff}
%%%%%%%%%%%%%%%%%%%%%%%%%%%%%%%%%%%%%%%%%%%%%%%%%%%%%%%
\subsection{A special class of laminates}\label{sec:WeffSimpl}
%%%%%%%%%%%%%%%%%%%%%%%%%%%%%%%%%%%%%%%%%%%%%%%%%%%%%%%
\subsubsection{General homogenisation result}
%%%%%%%%%%%%%%%%%%%%%%%%%%%%%%%%%%%%%%%%%%%%%%%%%%%%%%%

Let us now proceed with the determination of $\bm \theta$ in a special case. Here, we select a guess solution of the form ${\bm \theta} = a {\bf u}$ where the vector ${\bf u}$ was introduced in Eq.~\eqref{VecU}. Under this assumption, we find
\begin{equation}
	\begin{aligned}
	\widehat{\bf I} \, {\bf s}^{(\alpha)} = (1+a\tau^{(\alpha)}) {\bf A}^{(\alpha)} {\bf u}
	& = (1+a\tau^{(\alpha)}) 2 W_1^{(\alpha)} \left({\bf u} - R^{(\alpha)} {\bf v}\right) \\
	& = (1+a\tau^{(\alpha)}) 2 W_2^{(\alpha)} \left( {\bf u} / R^{(\alpha)} - {\bf v}\right)  , 
	\end{aligned}
	\label{TractionSimpl}
\end{equation}
where we have introduced
\begin{equation}
	{\bf v} = -\tilde I_5 \widehat{\bf I} \, \overline{\bf B}^{-1} {\bf u} = \widehat{\bf I} \, \overline{\bf B}^{-1} \overline{\bf F}^{-\text{T}} {\bf N} ,
	\label{VecV}
\end{equation}
and $R^{(\alpha)} = W_2^{(\alpha)}/W_1^{(\alpha)}$. Furthermore, using Eq.~\eqref{InvarHom}, we find
\begin{equation}
	\begin{aligned}
	&I_1^{(\alpha)}-3 = D + (1 + a\tau^{(\alpha)})^2 K , \\
	&I_2^{(\alpha)}-3 = E + (1 + a\tau^{(\alpha)})^2 L ,
	\end{aligned}
	\label{InvarSimpl}
\end{equation} 
where
\begin{equation}
	D = \tilde I_1-3-K, \quad E = \tilde I_2 - 3 - L ,
	\label{IdDE}
\end{equation}
and
\begin{equation}
	\begin{aligned}
	K = \tilde I_4 - \tilde I_5^{-1} ,\quad
	L = |\overline{\bf C}^{-1}{\bf N}|^2 \tilde I_5^{-1} - \tilde I_5 = \tilde I_2 - \tilde I_5 - \tilde I_5^{-1} (\tilde I_1 - \tilde I_4 ) .
	\end{aligned}
	\label{IdKL}
\end{equation}
Here, we have used the Cayley--Hamilton identity \eqref{CayleyHamilton} multiplied by $\overline{\bf C}^{-1}$ for the last equality.
The above quantities satisfy the relationships
\begin{equation}
	\begin{aligned}
	& K = |{\bf u}|^2 = \overline{\bf F}{\bf N}\cdot {\bf u} , \\
	& L = \tilde I_5 |\overline{\bf F}^{-1} {\bf u}|^2 = {-}\overline{\bf F}^{-\text{T}} {\bf N} \cdot \overline{\bf B}^{-1}{\bf u} = -{\bf u}\cdot {\bf v} = -\overline{\bf F}{\bf N} \cdot {\bf v} ,
	\end{aligned}
	\label{RelKL}
\end{equation}
in general.

The vectors \eqref{TractionSimpl} can be inserted in the traction continuity equation \eqref{TractionFinal} to find $a$. Unfortunately, it appears that the guess ${\bm \theta} = a{\bf u}$ does not solve the traction continuity equation whenever $R^{(1)} \neq R^{(2)}$, where the functions $R^{(\alpha)}$ were introduced in Eq.~\eqref{TractionSimpl}. We will address such a case later on (Section~\ref{sec:WeffMRPert}).

For now, we consider that the ratio $R^{(\alpha)}$ is the same in both layers. Since $R^{(1)}$ and $R^{(2)}$ are functions of different variables, we note that the only possibility for this to hold is that $R^{(\alpha)} = R$ is a constant. Solving this equation with respect to $(I_1^{(\alpha)},I_2^{(\alpha)})$, we conclude that
\begin{equation}
	\begin{aligned}
	W^{(\alpha)} &= \Upsilon^{(\alpha)} \left(I_1^{(\alpha)}-3+R (I_2^{(\alpha)}-3)\right) \\
	& = \Upsilon^{(\alpha)} \left(R \left( (I_1^{(\alpha)}-3)/R+ I_2^{(\alpha)}-3 \right)\right) ,
	\end{aligned}
	\label{WSimpl}
\end{equation}
where $\Upsilon^{(\alpha)}$ is an arbitrary function. Hence, the strain energy $W^{(\alpha)}$ in both phases is expressed as a function of the same linear combination of the invariants $I_1^{(\alpha)}$ and $I_2^{(\alpha)}$. This situation encompasses laminates whose layers are described by generalised neo-Hookean theories ($R=0$), or by similar theories defined in terms of the second strain invariant \citep{kuhl24}, for which $1/R=0$. If $\Upsilon^{(\alpha)}$ is a linear function, we obtain a family of Mooney--Rivlin layers for which the ratio $R$ between the Mooney parameters is the same in both phases.

Under the assumption $R^{(\alpha)} = R$, the traction continuity equation \eqref{TractionFinal} yields
\begin{equation}
	  a = \frac{W_1^{(1)} - W_1^{(2)}}{\nu^{(1)} W_1^{(2)} + \nu^{(2)} W_1^{(1)}} = \frac{W_2^{(1)} - W_2^{(2)}}{\nu^{(1)} W_2^{(2)} + \nu^{(2)} W_2^{(1)}} , 
	  \label{SolSimpl}
\end{equation}
where we have used Eq.~\eqref{TractionSimpl} and the definition of $\tau^{(\alpha)}$ in Eq.~\eqref{FdeBotton}. We note that the coefficients $W_1^{(\alpha)}$ and $W_2^{(\alpha)}$ are functions of the same linear combination of the invariants $I_1^{(\alpha)}$ and $I_2^{(\alpha)}$, see Eq.~\eqref{WSimpl}.
Given the expression \eqref{SolSimpl} of $a$ above, Eq.~\eqref{InvarSimpl} forms a nonlinear system of four equations in the four unknowns $I_1^{(\alpha)}$ and $I_2^{(\alpha)}$ where $\alpha = 1,2$, in general. As shown in the examples below, an exact solution might be found for special Mooney--Rivlin layers ($\Upsilon^{(\alpha)}$ linear), and an approximate solution can be obtained for generalised neo-Hookean theories ($R=0$). A similar calculation could be attempted for material models involving only the second strain invariant instead of the first one ($1/R=0$), see for instance \citet{kuhl24} for a description of such materials.

%%%%%%%%%%%%%%%%%%%%%%%%%%%%%%%%%%%%%%%%%%%%%%%%%%%%%%%
\subsubsection{Special Mooney--Rivlin laminates}
%%%%%%%%%%%%%%%%%%%%%%%%%%%%%%%%%%%%%%%%%%%%%%%%%%%%%%%

We consider now a subfamily of the special class of models \eqref{WSimpl} for which $\Upsilon^{(\alpha)}$ is a linear function. Thus, the coefficients $W_1^{(\alpha)}$ and $W_2^{(\alpha)}$ are constants, i.e., they do not depend on the invariants $I_1^{(\alpha)}$ and $I_2^{(\alpha)}$. Introducing the shear modulus $\mathcal{G}^{(\alpha)} = 2(W_1^{(\alpha)} + W_2^{(\alpha)})$ of the layer $\alpha$ and the ratio $R = W_2^{(\alpha)}/W_1^{(\alpha)}$, we can write
\begin{equation}
	W^{(\alpha)} = \tfrac12 (1-\delta) \mathcal{G}^{(\alpha)} ( I_1^{(\alpha)}-3 ) + \tfrac12 \delta \mathcal{G}^{(\alpha)} (I_2^{(\alpha)} - 3) ,
	\label{WMRSimpl}
\end{equation}
where $\delta = R / (1+R)$, or inversely, $R = \delta / (1-\delta)$.
This situation encompasses the neo-Hookean theory ($\delta = 0$), a similar theory based on the second strain invariant ($\delta=1$), as well as Mooney--Rivlin solids with $0< \delta < 1$.

Noting that
\begin{equation}
	1+a\tau^{(\alpha)} = \breve{\mathcal G}/\mathcal{G}^{(\alpha)} ,
	\quad
	\breve{\mathcal G} = \left(\frac{\nu^{(1)}}{\mathcal{G}^{(1)}} + \frac{\nu^{(2)}}{\mathcal{G}^{(2)}}\right)^{-1} ,
	\label{RelA}
\end{equation}
we can now evaluate the invariants \eqref{InvarSimpl} and the effective strain energy, $\widetilde{W} = \nu^{(1)} W^{(1)} + \nu^{(2)} W^{(2)}$. Doing so, we find
\begin{equation}
	\widetilde{W} =  \tfrac12 (1-\delta) \langle\mathcal{G}\rangle D + \tfrac12 \delta \langle\mathcal{G}\rangle E + \tfrac12 (1-\delta) \breve{\mathcal G} K + \tfrac12 \delta \breve{\mathcal G} L , 
	\label{WMRSimplEff}
\end{equation}
where $D$, $E$, $K$ and $L$ were introduced in Eqs.~\eqref{IdDE}-\eqref{IdKL}, and $\langle\mathcal G\rangle = \nu^{(1)}\mathcal{G}^{(1)} + \nu^{(2)}\mathcal{G}^{(2)}$. In other words, $\langle\mathcal G\rangle$ is the linear average of the shear moduli $\mathcal{G}^{(\alpha)}$, whereas $\breve{\mathcal G} = \langle\mathcal{G}^{-1}\rangle^{-1}$ is their harmonic average. Thus, according to the mean inequality chain, we have $\langle \mathcal{G} \rangle \geq \breve{\mathcal G}$. For $\delta =0$, we recover the classical result obtained for neo-Hookean layers.

According to the above result, the effective strain energy $\widetilde{W}$ depends on the invariants \eqref{InvarMacro} through the quantities $D$, $E$, $K$ and $L$. Incidentally, we note that the effective strain energy function can always be decomposed as a sum of the form \citep{horgan20}
\begin{equation}
	\widetilde{W} =  \widetilde{W}_\text{iso} + \widetilde{W}_\text{an} ,
	\label{WDec}
\end{equation}
where 
\begin{equation}
	\widetilde{W}_\text{iso} = \left. \left(\widetilde{W}|_{(D,E)=(\tilde I_1-3,\tilde I_2-3)} - \widetilde{W}|_{(D,E)=(0,0)} \right) \right|_{(K,L)=(0,0)} .
\end{equation}
The energy $\widetilde{W}_\text{iso}$ depends on $(\tilde I_1,\tilde I_2)$ only (note that the above choice is not unique), whereas $\widetilde{W}_\text{an} = \widetilde{W} - \widetilde{W}_\text{iso}$ might depend on $\tilde I_4$ and $\tilde I_5$ as well. Here, using Eq.~\eqref{WMRSimplEff}, we have
\begin{equation}
	\begin{aligned}
		\widetilde{W}_\text{iso} &= \tfrac12 \langle\mathcal{G}\rangle \left( (1-\delta)  (\tilde I_1-3) + \delta (\tilde I_2-3) \right) ,\\
		\widetilde{W}_\text{an} &= \tfrac12 (\breve{\mathcal G} - \langle\mathcal{G}\rangle) \left( (1-\delta) (\tilde I_4 - \tilde I_5^{-1})  %\right.\\
	%& \left. \qquad \quad \qquad\phantom{0}
	 + \delta (\tilde I_2 - \tilde I_5 - \tilde I_5^{-1} (\tilde I_1 - \tilde I_4 ))\right) .
	\end{aligned}
	\label{WDecMR}
\end{equation}
We observe that for neo-Hookean laminates ($\delta=0$), the anisotropic contribution to the effective strain energy $\widetilde{W}_\text{an}$ depends on $(\tilde I_4,\tilde I_5)$ only. However, it is worth noting that this is no longer true when $\delta \neq 0$, and that the present definition \eqref{InvarMacro} of $\tilde I_5$ differs from common choices found in the literature.

If $\widetilde{W}$ is expressed in terms of the new scalar functions introduced in Eqs.~\eqref{IdDE}-\eqref{IdKL}, then using differentiation rules, we find
\begin{equation}
	\begin{aligned}
	&\widetilde{W}_1 = \widetilde{W}_D - \tilde I_5^{-1} (\widetilde{W}_L - \widetilde{W}_E) , \\
	&\widetilde{W}_2 = \widetilde{W}_L, \\
	&\widetilde{W}_4 = \widetilde{W}_K - \widetilde{W}_D + \tilde I_5^{-1} (\widetilde{W}_L - \widetilde{W}_E) , \\
	&\widetilde{W}_5 = \tilde I_5^{-2} (\widetilde{W}_K - \widetilde{W}_D) + ((\tilde I_1 - \tilde I_4)\tilde I_5^{-2} - 1) (\widetilde{W}_L-\widetilde{W}_E) ,
	\end{aligned}
	\label{RelW}
\end{equation}
see the computations in the Appendix~\ref{app:Acoust}.
Again, the coefficients $\widetilde{W}_\square = \partial \widetilde{W}/\partial \square$ with $\square = D, E, K, L$ are partial derivatives of $\widetilde{W}$. The effective first Piola--Kirchhoff stress tensor $\overline{\bf P} = \overline{\bf T} \overline{\bf F}^{-\text{T}}$ can be deduced from these relationships using Eq.~\eqref{StressMacro}, which yields the Cauchy stress tensor
\begin{equation}
	\begin{aligned}
	\overline{\bf T} & =  - \overline{p}\, {\bf I} + 2 \widetilde{W}_D \overline{\bf B} - 2 \widetilde{W}_L \overline{\bf B}^{-1} \\
	& + 2 (\widetilde{W}_K - \widetilde{W}_D) \left( \tilde I_4 {\bf n} \otimes {\bf n} - \tilde I_5^{-1} {\bf m} \otimes {\bf m} \right) \\
	& \, {-} \, 2 (\widetilde{W}_L - \widetilde{W}_E) \left(\tilde I_5^{-1}\overline{\bf B} - \tilde I_4 \tilde I_5^{-1} {\bf n} \otimes {\bf n} %\right.\\
	%&\qquad\qquad\qquad \left. \phantom{\overline{\bf B}} 
	+ ( (\tilde I_1 - \tilde I_4)\tilde I_5^{-1} - \tilde I_5) {\bf m} \otimes {\bf m}\right) . 
	\end{aligned}
	\label{PMR}
\end{equation}
The present constitutive law is consistent with the one obtained in the neo-Hookean limit ($\delta\to 0$), where $2 \widetilde{W}_D \to \langle\mathcal G\rangle$ and $2 \widetilde{W}_K \to \breve{\mathcal G}$ while $\widetilde{W}_E \to 0$ and $\widetilde{W}_L \to 0$, see for instance \citet{conroy24}.

%%%%%%%%%%%%%%%%%%%%%%%%%%%%%%%%%%%%%%%%%%%%%%%%%%%%%%%
\subsubsection{Generalised neo-Hookean laminates}
%%%%%%%%%%%%%%%%%%%%%%%%%%%%%%%%%%%%%%%%%%%%%%%%%%%%%%%

In generalised neo-Hookean layers, the strain energy $W^{(\alpha)}$ is an arbitrary smooth function of the first strain invariant only, $I_1^{(\alpha)}$. Thus, this class of constitutive models is a subfamily of the special class of models \eqref{WSimpl} for which $R=0$. Typical examples of such constitutive laws are the Yeoh, Fung--Demiray and Gent theories. For the two-term Yeoh strain energy function, \citet{berjamin25} solved the system of coupled equations \eqref{InvarSimpl} in an approximate fashion, and the expression of the effective strain energy $\widetilde W$ could be established. To do so, the Yeoh strain energy function was viewed as a small perturbation of the neo-Hookean strain energy, for which the effective strain energy is known.

Any smooth strain energy function $W^{(\alpha)}$ of this kind can be written as a polynomial Yeoh-type energy using Taylor series approximations of an arbitrary order in terms of $(I_1^{(\alpha)}-3)$. However, at quadratic orders of $(I_1^{(\alpha)}-3)$, most constitutive models of generalised neo-Hookean type cannot be distinguished from the two-term Yeoh theory \citep{berjamin25}. Hence, to analyse generalised neo-Hookean laminates using this type of approximation, it is necessary to use higher-order expansions of the strain energy function.

If we restrict ourselves to the third order of the Taylor series expansion of $W^{(\alpha)}$ at $I_1^{(\alpha)}=3$, then we end up with the three-term Yeoh energy function
\begin{equation}
	\begin{aligned}
	W^{(\alpha)} &= \tfrac12 \mathcal{G}^{(\alpha)} (I_1^{(\alpha)} - 3) + \tfrac14 b^{(\alpha)} \mathcal{G}^{(\alpha)} (I_1^{(\alpha)} - 3)^2 \\
		&\quad + \tfrac1{12} d^{(\alpha)} \mathcal{G}^{(\alpha)} (I_1^{(\alpha)} - 3)^3 ,
	\end{aligned}
	\label{WYeoh3}
\end{equation}
where the coefficients $b^{(\alpha)}$ and $d^{(\alpha)}$ are dimensionless. For instance, the Gent and Fung--Demiray energy functions defined by
\begin{equation}
	\begin{aligned}
	&W_{\text{G}}^{(\alpha)} = -\frac{\mathcal{G}^{(\alpha)}}{2 b^{(\alpha)}} \ln \left(1 - b^{(\alpha)} (I_1^{(\alpha)}-3)\right) , \\
	&W_{\text{FD}}^{(\alpha)} = \frac{\mathcal{G}^{(\alpha)}}{2 b^{(\alpha)}} \left( \text{e}^{b^{(\alpha)} (I_1^{(\alpha)}-3)} - 1\right) ,
	\end{aligned}
	\label{WGent}
\end{equation}
are consistent with Eq.~\eqref{WYeoh3} in the vicinity of $I_1^{(\alpha)}=3$, with the parameters $d^{(\alpha)} = k^{(\alpha)} (b^{(\alpha)})^2$ and $k^{(\alpha)}=\lbrace 2, 1\rbrace$, respectively. Therefore, at this order, they differ only in the coefficient $d^{(\alpha)}$ accounting for third-order terms. The two-term Yeoh model is recovered for $d^{(\alpha)} = 0$.

For the two-term Yeoh energy ($d^{(\alpha)} = 0$), the nonlinear algebraic system of equations \eqref{InvarSimpl} where $a$ is dependent on $I_1^{(1)}$ and $I_1^{(2)}$ can be solved in an approximate fashion by linearisation of the equations and of their solution using Taylor series expansions valid for $b^{(\alpha)} \ll 1$. At leading order in $b^{(1)}$ and $b^{(2)}$, we obtain a polynomial expression for the energy, which takes the form \citep{berjamin25}
\begin{equation}
	\begin{aligned}
	&\widetilde{W} = \tfrac12 \langle\mathcal G\rangle D + \tfrac12 \breve{\mathcal G} K \\
	&\qquad + \tfrac14 \langle \mathcal{G} b \rangle D^2 + \tfrac12 \breve{\mathcal{G}}^2 \langle \mathcal{G}^{-1} b \rangle  DK + \tfrac14 \breve{\mathcal{G}}^4 \langle \mathcal{G}^{-3} b \rangle K^2 ,
	\end{aligned}
	\label{WYeoh}
\end{equation}
where the same notation as earlier is used, see Eq.~\eqref{WMRSimplEff}. The averaged quantities above are evaluated as follows,
\begin{equation}
	\langle \mathcal{G}^n b \rangle = \nu^{(1)} (\mathcal{G}^{(1)})^n b^{(1)} + \nu^{(2)} (\mathcal{G}^{(2)})^n b^{(2)} ,
	\label{AverNotation}
\end{equation}
for $n$ integer. This expression of $\widetilde{W}$ can be obtained directly using symbolic calculus software, or by exploiting relevant limiting cases.

We could follow the same approach for the study of generalised neo-Hookean laminates described by strain energy functions of the form \eqref{WYeoh3}, where $d^{(\alpha)}$ is not necessarily equal to zero. For the sake of simplicity, we begin with investigating the influence of cubic terms $\propto (I_1^{(\alpha)}-3)^3$ on the effective strain energy, and we set $b^{(\alpha)} = 0$ in Eq.~\eqref{WYeoh3}, while $d^{(\alpha)}$ is a small dimensionless parameter. We rely again on a perturbation approach to solve Eq.~\eqref{InvarSimpl}, and we arrive at an expression of the effective strain energy which is of a similar form to Eq.~\eqref{WYeoh}. The only difference is that the terms proportional to $D^i K^j$ with $i+j=2$ are now replaced by similar terms where $i+j=3$. We follow the same steps as earlier to determine the coefficients in this sum by means of a symbolic calculus software. Finally, we arrive at
\begin{equation}
	\begin{aligned}
	&\widetilde{W} = \tfrac12 \langle\mathcal G\rangle D + \tfrac12 \breve{\mathcal G} K \\
	&\qquad + \tfrac1{12} \langle\mathcal{G}d\rangle D^3 + \tfrac1{4} \breve{\mathcal G}^2 \langle\mathcal{G}^{-1}d\rangle D^2K \\
	&\qquad + \tfrac1{4} \breve{\mathcal G}^4 \langle\mathcal{G}^{-3}d\rangle DK^2 + \tfrac1{12} \breve{\mathcal G}^6 \langle\mathcal{G}^{-5}d\rangle K^3 ,
	\end{aligned}
	\label{WThird}
\end{equation}
where similar notation is used as for Eq.~\eqref{WYeoh}.

Based on the above homogenisation results, we conclude that a laminate made of soft materials described by the energy function \eqref{WYeoh3} can be represented by the following effective strain energy function (verified using symbolic calculus software),
\begin{equation}
	\begin{aligned}
	& \widetilde{W} = \tfrac12 \langle\mathcal G\rangle D + \tfrac12 \breve{\mathcal G} K \\
	&\qquad + \tfrac14 \langle\mathcal{G} b\rangle D^2 + \tfrac12 \breve{\mathcal{G}}^2 \langle\mathcal{G}^{-1} b\rangle DK + \tfrac14 \breve{\mathcal{G}}^4 \langle\mathcal{G}^{-3} b\rangle K^2 \\
	&\qquad + \tfrac1{12} \langle\mathcal{G}d\rangle D^3 + \tfrac1{4} \breve{\mathcal G}^2 \langle\mathcal{G}^{-1}d\rangle D^2K \\
	&\qquad + \tfrac1{4} \breve{\mathcal G}^4 \langle\mathcal{G}^{-3}d\rangle DK^2 + \tfrac1{12} \breve{\mathcal G}^6 \langle\mathcal{G}^{-5}d\rangle K^3 ,
	\end{aligned}
	\label{WGenNH}
\end{equation}
at leading order in the small parameters $b^{(\alpha)}$ and $d^{(\alpha)}$. We emphasise that correcting terms \eqref{WGenNHCorr} need to be added to Eq.~\eqref{WGenNH} if $d^{(\alpha)} = k^{(\alpha)} b^{(\alpha) 2}$ where the parameters $b^{(\alpha)}$ are assumed small, as specified subsequently.
In passing, we observe that the homogeneous limit $\nu^{(2)} \to 0$ yields $\widetilde{W} \to W^{(1)}$ with the three-term expression \eqref{WYeoh3}, and a similar observation can be made as $\nu^{(2)} \to 1$, or as $W^{(1)} \to W^{(2)}$. In this limit, the effective strain energy is only function of $D+K=\tilde I_1-3$.

According to Eq.~\eqref{WGenNH}, the effective strain energy $\widetilde{W}$ depends on the invariants \eqref{InvarMacro} through the scalar quantities $D$ and $K$ only, see their definition in Eqs.~\eqref{IdDE}-\eqref{IdKL}. The effective strain energy can be decomposed into the same form as Eq.~\eqref{WDec}, using for instance
\begin{equation}
	\widetilde{W}_\text{iso} = \frac{\langle\mathcal G\rangle}{2} \left( \tilde I_1 - 3 + \tfrac12 \bar b (\tilde I_1 - 3)^2 + \tfrac1{6} \bar d (\tilde I_1 - 3)^3 \right) ,
	\label{WDecGenNH}
\end{equation}
where $\bar b = \langle\mathcal{G} b\rangle/\langle\mathcal G\rangle$ and $\bar d = \langle\mathcal{G}d\rangle / \langle\mathcal G\rangle$. This way, the energy contribution $\widetilde{W}_\text{iso}$ is of the same type as the strain energy in the layers, see Eq.~\eqref{WYeoh3}.
The effective first Piola--Kirchhoff stress tensor $\overline{\bf P} = \overline{\bf T} \overline{\bf F}^{-\text{T}}$ can be deduced from the Cauchy stress tensor \eqref{PMR} with $\widetilde{W}_L = \widetilde{W}_E = 0$. The present constitutive law is consistent with the one obtained in the neo-Hookean limit ($b^{(\alpha)}, d^{(\alpha)} \to 0$).

One temporary shortcoming of the derivation leading to Eq.~\eqref{WGenNH} is that we have treated the quadratic terms \eqref{WYeoh} separately from the cubic terms \eqref{WThird}. Of course, this is perfectly valid when considering Yeoh materials for which the second-order coefficients $b^{(\alpha)}$ are independent from the third-order coefficients, $d^{(\alpha)}$. However, this derivation is questionable when considering generalised neo-Hookean materials \eqref{WGenNH} for which the relationship $d^{(\alpha)} = k^{(\alpha)} b^{(\alpha) 2}$ holds (i.e., the small parameters $b^{(\alpha)}$ and $d^{(\alpha)}$ are not independent). In fact, the terms of the same order as $d^{(\alpha)}$ in Eq.~\eqref{WGenNH} are now of order $b^{(\alpha) 2}$. For consistency, we therefore need to expand the invariants and the energy up to quadratic orders of the small parameters $b^{(\alpha)}$. Using symbolic calculus software (cf. supplementary material), we find that the following correction needs to be added to the expression of the effective energy,
\begin{equation}
	\begin{aligned}
	 \widetilde{W} &= \widetilde{W}_\text{YG} \\
	&\quad {-} \frac{\nu^{(1)}\nu^{(2)} \breve{\mathcal G}^3}{\mathcal{G}^{(1)}\mathcal{G}^{(2)}} \left( (b^{(2)}-b^{(1)}) D + \left( \frac{b^{(2)}}{\mathcal{G}^{(2) 2}} - \frac{b^{(1)}}{\mathcal{G}^{(1) 2}} \right) \breve{\mathcal G}^2 K \right)^2 K ,
	\end{aligned}
	\label{WGenNHCorr}
\end{equation}
where $\widetilde{W}_\text{YG}$ is the effective energy \eqref{WGenNH} with $d$ replaced by $k b^{2}$. We will see in Sections~\ref{sec:Valid}-\ref{sec:Insta} that the above correction can sometimes be neglected in applications, supporting the use of Eq.~\eqref{WGenNH}. Furthermore, we note that this correction vanishes in the homogeneous limit, as well as in the limit of two-term Yeoh materials \eqref{WYeoh}, at the relevant order of approximation. Despite its asymptotic nature, the present analytical result obtained through homogenisation contrasts with the \emph{ad hoc} approach used by \citet{yao24} to describe Gent laminates, beyond the potential errors in some equations therein \citep{berjamin25}.\footnote{In the macroscopic theory proposed by \citet{yao24}, the effective energy (23) depends on phase-related deformations through Eqs.~(24)-(25), whereas it is a function of the macroscopic deformation only, by definition. Furthermore, the effective shear moduli in Eqs. (24) and (27) therein should be the arithmetic and harmonic means of the layers' shear moduli. \label{foot:yao}}

%%%%%%%%%%%%%%%%%%%%%%%%%%%%%%%%%%%%%%%%%%%%%%%%%%%%%%%
\subsection{Mooney--Rivlin laminates}\label{sec:WeffMRPert}
%%%%%%%%%%%%%%%%%%%%%%%%%%%%%%%%%%%%%%%%%%%%%%%%%%%%%%%

In the previous section, we have studied special Mooney--Rivlin layers whose constitutive law belongs to the class of models \eqref{WSimpl}. We now use a similar asymptotic approach as above to study general Mooney--Rivlin layers. Thus, we set
\begin{equation}
	W^{(\alpha)} = \tfrac12 (1 - \delta^{(\alpha)}) \mathcal{G}^{(\alpha)} (I_1^{(\alpha)}-3) + \tfrac12 \delta^{(\alpha)} \mathcal{G}^{(\alpha)} (I_2^{(\alpha)}-3) ,
	\label{WMRPhase}
\end{equation}
where the parameters $\delta^{(\alpha)} = \delta + \epsilon\zeta^{(\alpha)}$ are dimensionless, and $\zeta^{(2)}-\zeta^{(1)} = 1$. This way, the ratio $R^{(\alpha)}$ introduced in Eq.~\eqref{TractionSimpl} equals $\delta^{(\alpha)}/(1 - \delta^{(\alpha)})$ in each phase, and we have $\delta^{(2)}-\delta^{(1)}=\epsilon$. The special case \eqref{WMRSimpl} addressed earlier is recovered when $R^{(\alpha)} = R$ is a given constant, or equivalently, when the parameters $\delta^{(\alpha)}=\delta$ are the same in both phases (i.e., $\epsilon = 0$).

For the present model, the traction continuity equation \eqref{TractionFinal} seems difficult to solve in an exact fashion, and the trial expression ${\bm \theta} = a {\bf u}$ is not a solution. We rely again on an approximate method of resolution which takes the form of a perturbation approach based on the parameter $\epsilon$, which is assumed to be infinitesimally small. In other words, we consider Mooney--Rivlin layers which are viewed as a small perturbation of the special case \eqref{WMRSimpl}. We set ${\bm\theta} = {\bm\theta}_{0} + {\bm\theta}_{1}$, where ${\bm\theta}_{1}$ is linear in $\epsilon$, and higher polynomial orders of $\epsilon$ are neglected. At leading order ($\epsilon = 0$), we expect that the solution of the traction continuity equation is the one obtained earlier, i.e., ${\bm\theta}_{0} = a_0 {\bf u}$ where $a_0 = a$ is given by Eq.~\eqref{SolSimpl} with $2W^{(\alpha)}_1 = (1-\delta)\mathcal{G}^{(\alpha)}$ and $2W^{(\alpha)}_2 = \delta \mathcal{G}^{(\alpha)}$.

To estimate the first-order corrections, we insert the above expansion in the expression of the projected tractions \eqref{TractionProj}, which are then linearised with respect to $\epsilon$ assuming a small value of this parameter. Doing so, we find
\begin{equation}
	\begin{aligned}
	\widehat{\bf I} \, {\bf s}^{(\alpha)} &= \breve{\mathcal{G}} \left( {\bf u} - \delta^{(\alpha)} ({\bf u} + {\bf v})\right) %\\
		%&\quad 
		+ \tau^{(\alpha)} \mathcal{G}^{(\alpha)} \left( (1-\delta) {\bf I} + \delta \tilde I_5 \widehat{\bf I} \, \overline{\bf B}^{-1} \right) {\bm \theta}_1 , 
	\end{aligned}
	\label{TractionPert}
\end{equation}
where we have used Eq.~\eqref{VecU} and the relationship \eqref{RelA} satisfied by $a_0$. Based on this formula and on the traction continuity equation \eqref{TractionFinal}, we observe that the determination of ${\bm \theta}_1$ can hardly be achieved if $\delta=1$ since the projection tensor $\widehat{\bf I}$ is singular. However, this calculation is greatly simplified if we set $\delta=0$, that is, if we restrict our attention to Mooney--Rivlin layers which are viewed as a small perturbation of neo-Hookean layers.

For $\delta=0$, we can easily determine the first-order correction ${\bm \theta}_1$ above using the continuity of tractions \eqref{TractionFinal} at the layer interfaces. The resulting vector ${\bm \theta}_1 = a_1 ({\bf u} + {\bf v})$ is determined according to the formula
\begin{equation}
	a_1 = \frac{\epsilon \breve{\mathcal{G}}}{\nu^{(1)} \mathcal{G}^{(2)} + \nu^{(2)} \mathcal{G}^{(1)}} = \frac{\epsilon \breve{\mathcal{G}}^2}{\mathcal{G}^{(1)}\mathcal{G}^{(2)}} ,
	\label{APert}
\end{equation}
see the definition of $\breve{\mathcal{G}}$ in Eq.~\eqref{RelA}, and this expression can now be used to determine first-order corrections of the invariants \eqref{InvarHom}. Upon linearisation with respect to $\epsilon$, the first invariant becomes
\begin{equation}
	I_1^{(\alpha)} = I_{1,0}^{(\alpha)} + I_{1,1}^{(\alpha)}, \quad I_{1,1}^{(\alpha)} = 2 a_1 \tau^{(\alpha)} (\breve{\mathcal G} / \mathcal{G}^{(\alpha)}) (K-L) ,
	\label{InvarPert} 
\end{equation}
where we have used the identities \eqref{RelKL} and the relationship \eqref{RelA} satisfied by $a_0$. Here, the leading-order term $I^{(\alpha)}_{1,0}$ in the expression of the first invariant is obtained from Eqs.~\eqref{InvarSimpl} and \eqref{RelA} with $a=a_{0}$.

Up to the first order of the small parameter $\epsilon$, the expression of the effective strain energy function $\widetilde{W}$ deduced from Eqs.~\eqref{WMRPhase}-\eqref{InvarPert} is as follows,
\begin{equation}
	\widetilde{W} = \tfrac12 \langle\mathcal{G}\rangle D + \tfrac12 \breve{\mathcal G} K + \tfrac12 \sum_{\alpha=1}^2 \nu^{(\alpha)} \delta^{(\alpha)} \mathcal{G}^{(\alpha)} (I_{2,0}^{(\alpha)} - I_{1,0}^{(\alpha)})  ,
\end{equation}
where the first two terms are identical to the first terms of Eq.~\eqref{WMRSimplEff} with $\delta=0$, and the following terms are linear in $\epsilon = \delta^{(\alpha)}/\zeta^{(\alpha)}$. In a similar fashion to the calculation of $I_{1,0}^{(\alpha)}$, the leading-order term $I^{(\alpha)}_{2,0}$ in the expression of the second invariant is obtained from Eqs.~\eqref{InvarSimpl} and \eqref{RelA} with $a=a_{0}$. Thus, based on these expressions, we can rewrite the above formula using averages in the form
\begin{equation}
	\begin{aligned}
	\widetilde{W} &= \tfrac12 \langle \mathcal{G} (1-\delta)\rangle D + \tfrac12 \langle \mathcal{G} \delta\rangle E \\
	&\quad + \tfrac12 \breve{\mathcal G}^2 \langle \mathcal{G}^{-1} (1-\delta) \rangle K + \tfrac12 \breve{\mathcal G}^2 \langle \mathcal{G}^{-1} \delta \rangle L ,
	\end{aligned}
	\label{WMREff}
\end{equation}
with similar notation as earlier. We note that this expression is consistent with the Mooney--Rivlin energy function in the homogeneous limit ($\nu^{(2)} \to 0$).

Finally, up to modified coefficients, we obtain an effective energy which is similar to the one obtained for special Mooney--Rivlin layers, see Eq.~\eqref{WMRSimplEff}, which is recovered for $\delta^{(1)}=\delta^{(2)}$. Nevertheless, it is worth pointing out that the above expression was obtained asymptotically for Mooney--Rivlin solids which are viewed as a small perturbation of neo-Hookean solids ($\delta=0$, $\epsilon \ll 1$). Therefore, this result is likely to be amended in other cases, e.g. for $\delta\neq 0$ or at higher polynomial orders of the small parameter $\epsilon$, for which the explicit evaluation of the effective strain energy function $\widetilde{W}$ becomes increasingly difficult within the present homogenisation framework. A validation of the asymptotic homogenised strain energies \eqref{WGenNH} and \eqref{WMREff} is carried out hereinafter.

%%%%%%%%%%%%%%%%%%%%%%%%%%%%%%%%%%%%%%%%%%%%%%%%%%%%%%%
\section{Validation}\label{sec:Valid}
%%%%%%%%%%%%%%%%%%%%%%%%%%%%%%%%%%%%%%%%%%%%%%%%%%%%%%%

The effective energy \eqref{WMRSimplEff} obtained for special Mooney--Rivlin laminates is an exact analytical result. It encompasses neo-Hookean laminates as a special case ($\delta=0$), and similar models of layered media for which the strain energy is linear in the second strain invariant ($\delta=1$). Therefore, the accuracy of this homogenisation result is only restricted by the assumptions underpinning its derivation, namely the constitutive behaviour of the layers, the perfect bonding at the layer interfaces, and the assumed deformations (which are periodic and homogeneous in each phase).

Compared to Eq.~\eqref{WMRSimplEff}, the effective energy functions \eqref{WGenNH} and \eqref{WMREff} obtained for three-term Yeoh layers and for Mooney--Rivlin layers have a different status, in that they were obtained within a certain asymptotic limit (Table~\ref{tab:Expr}). In what follows, we investigate the validity of our asymptotic homogenisation results by comparing them to reference solutions, either exact analytical expressions obtained in special cases, or numerical solutions of the constrained minimisation problem \eqref{WMiniDeBotton}. Here, we assess the potential errors made in the estimation of the macroscopic energy $\widetilde{W}$. A similar validation could be performed by measuring the errors in the stress components \eqref{StressMacro} or in the incremental moduli, see \citet{lopezpamies09} for the determination of these quantities based on a solution to the minimisation problem \eqref{WMiniDeBotton}. A validation of this kind is carried out for specific orientations in Section~\ref{subsec:Wave}.

%%%%%%%%%%%%%%%%%%%%%%%%%%%%%%%%%%%%%%%%%%%%%%%%%%%%%%%
\subsection{Homogeneous deformations}
%%%%%%%%%%%%%%%%%%%%%%%%%%%%%%%%%%%%%%%%%%%%%%%%%%%%%%%

Consider homogeneous deformations such that ${\bf F}^{(\alpha)} = \overline{\bf F}$ in both phases. Under this assumption, the macroscopic elastic energy can be estimated directly according to Eq.~\eqref{WMiniDeBotton}. Furthermore, the layers need to satisfy the continuity equation \eqref{RepDef}, from where we deduce ${\bm\theta} = {\bf 0}$. The material is in equilibrium provided that the projected tractions \eqref{TractionProj} are continuous across the layer interfaces. This condition requires that $\bf u$ belongs to the kernel of ${\bf A}^{(2)} - {\bf A}^{(1)}$, which can be viewed as a restriction on the macroscopic deformations or on the constitutive response of the layers.

To avoid making assumptions on the behaviour of the layers, we consider all the potential macroscopic deformations such that ${\bf u}={\bf 0}$, i.e.,
\begin{equation}
	\overline{\bf F} {\bf N} = \tilde{I}_5^{-1} \overline{\bf F}^{-\text{T}} {\bf N} .
\end{equation}
Using Eqs.~\eqref{IdDE}-\eqref{IdKL}, we deduce from this restriction that $K=L=0$, and that $D= \tilde I_1-3$, $E = \tilde I_2-3$. Thus, for the present class of homogeneous deformations, it can be observed that the effective strain energy functions \eqref{WGenNH} and \eqref{WMREff} are exact, in that they provide the same value of the average energy as Eq.~\eqref{WMiniDeBotton}. For the three-term Yeoh model \eqref{WGenNH}, the corrections \eqref{WGenNHCorr} have no incidence on this conclusion since $K=0$.

%%%%%%%%%%%%%%%%%%%%%%%%%%%%%%%%%%%%%%%%%%%%%%%%%%%%%%%
\subsection{Uniaxial extension and simple shear}
%%%%%%%%%%%%%%%%%%%%%%%%%%%%%%%%%%%%%%%%%%%%%%%%%%%%%%%

\begin{figure}
    \centering
	\includegraphics{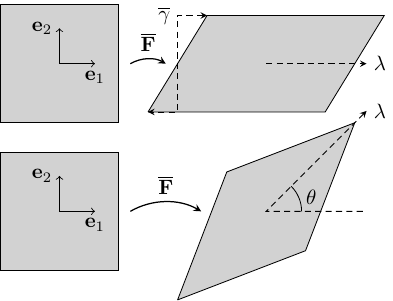}
	\caption{Effect of the extension-shear deformation \eqref{FValid} (top) and of the pure shear deformation \eqref{FPureShear} (bottom) on a square domain aligned with the basis vectors ${\bf e}_1$, ${\bf e}_2$. In the upper part of the figure, a stretch ratio $\lambda=1.5$ is applied along ${\bf e}_1$, combined with a shearing ratio $\overline{\gamma} = 0.5$. Below, a stretch ratio $\lambda=1.5$ is applied along the direction of polar angle $\theta=\pi/4$. \label{fig:Stretch}}
\end{figure}

Let us consider an infinite layered medium with alternating layers. We assume that each phase undergoes a simple shearing motion combined with uniform extension (Figure~\ref{fig:Stretch}). With respect to a suitable basis, this motion can be described by the deformation gradient tensors
\begin{equation}
	{\bf F}^{(\alpha)} = \begin{bmatrix}
		\lambda & \gamma^{(\alpha)} & 0 \\
		0 & \lambda^{-1/2} & 0 \\
		0 & 0 & \lambda^{-1/2}
	\end{bmatrix} ,
	\label{FValid}
\end{equation}
where $\lambda > 0$ is the stretching ratio, and $\gamma^{(\alpha)}$ represents the amount of shear in each phase. Based on the above result concerning certain types of homogeneous deformations, it can be shown that the proposed homogenised theories are exact in uniaxial extension along any direction ($\gamma^{(\alpha)}=0$). Hence, we focus here on shearing motions for which the shear strains can take arbitrary values.

The strain invariants \eqref{InvarMicro} in each phase are given by
\begin{equation}
	I_1^{(\alpha)} = \lambda^2 + 2\lambda^{-1} + \gamma^{(\alpha) 2}, \quad
	I_2^{(\alpha)} = \lambda^{-2} + 2\lambda + \lambda^{-1} \gamma^{(\alpha) 2} .
	\label{InvarValid}
\end{equation}
The continuity requirement \eqref{RepDef} yields ${\bm\theta} = [\theta_1,0,0]$ where $\theta_1 = \gamma^{(2)} - \gamma^{(1)}$, and ${\bf N} = {\bf e}_2$. The material is in equilibrium provided that the traction continuity equation \eqref{TractionFinal} is satisfied, i.e.,
\begin{equation}
	2 (W^{(1)}_1 + \lambda^{-3} W^{(1)}_2) \gamma^{(1)} = 2 (W^{(2)}_1 + \lambda^{-3} W^{(2)}_2) \gamma^{(2)} ,
	\label{RelShear}
\end{equation}
where we have used Eq.~\eqref{FdeBotton}. In general, Eq.~\eqref{RelShear} is nonlinear with respect to the shearing ratios, $\gamma^{(\alpha)}$, which are needed to evaluate the macroscopic strain energy \eqref{WMiniDeBotton}.

Next, we evaluate the effective quantities involved in the analytical estimates of the macroscopic strain energy. The deformation gradient $\overline{\bf F}$ is of the same form as ${\bf F}^{(\alpha)}$, with the average amount of shear $\overline{\gamma} = \nu^{(1)}\gamma^{(1)} + \nu^{(2)}\gamma^{(2)}$ instead of $\gamma^{(\alpha)}$. Consequently, the macroscopic invariants $\tilde I_1$ and $\tilde I_2$ defined in Eq.~\eqref{InvarMacro} are of the same form as their counterpart from Eq.~\eqref{InvarValid}, with $\gamma^{(\alpha)}$ replaced by $\overline{\gamma}$, whereas the value taken by the invariants $\tilde I_4$ and $\tilde I_5$ depends on the lamination direction. For ${\bf N} = {\bf e}_2$, we find
\begin{equation}
	D = \lambda^2 + 2\lambda^{-1} - 3, \quad
	E = \lambda^{-2} + 2\lambda - 3 ,
	\quad	
	K = \overline{\gamma}^2 = \lambda L, 
	\label{EffValid}
\end{equation}
using Eqs.~\eqref{IdDE}-\eqref{IdKL}. These expressions can be used to evaluate the homogenised strain energies.

\paragraph{Yeoh laminates.} To keep calculations tractable, we restrict this validation to materials described by the two-term Yeoh energy function \eqref{WYeoh3} under simple shear (i.e., $k^{(\alpha)} = 0$ in both phases and $\lambda=1$). We solve the above equations in an asymptotic fashion by using Taylor series expansions of the shear strain components in terms of the parameters of nonlinearity ($b^{(\alpha)} \simeq 0$). Doing so, we find
\begin{equation}
	\gamma^{(\alpha)} \simeq \frac{\breve{\mathcal G} \overline{\gamma}}{\mathcal{G}^{(\alpha)}}
	 + \frac{\nu^{(\beta)} \breve{\mathcal G}^4 \overline{\gamma}^3}{\mathcal{G}^{(1)} \mathcal{G}^{(2)}} \left( \frac{b^{(\beta)}}{\mathcal{G}^{(\beta)2}}  - \frac{b^{(\alpha)}}{\mathcal{G}^{(\alpha)2}}  \right),
	 \label{ShearYeohValid}
\end{equation}
where $\beta \neq \alpha$.
Substitution into Eq.~\eqref{WMiniDeBotton} together with \eqref{InvarValid} yields
\begin{equation}
	\widetilde{W} \simeq \tfrac12 \breve{\mathcal G} \overline{\gamma}^2 + \tfrac14 \breve{\mathcal G}^4 \langle {\mathcal G}^{-3} b \rangle \overline{\gamma}^4 ,
\end{equation}
at leading order in $b^{(1)}$ and $b^{(2)}$. Comparison with Eqs.~\eqref{WYeoh} and \eqref{EffValid} shows that the effective theory is at least asymptotically correct under simple shear along the lamination direction. This conclusion was expected, by construction. Furthermore, at the present order of approximation, the corrections \eqref{WGenNHCorr} have no incidence on this conclusion.

\paragraph{Mooney--Rivlin laminates.} For layers described by the Mooney--Rivlin energy function \eqref{WMRPhase}, the system formed by Eq.~\eqref{RelShear} and by the definition of $\overline{\gamma}$ can be solved analytically. Using similar notations as above, we arrive at
\begin{equation}
	\gamma^{(\alpha)} \simeq \frac{\breve{\mathcal G} \overline{\gamma}}{\mathcal{G}^{(\alpha)}} +  \frac{(-1)^\beta (\lambda^3-1)}{\delta^{(1)}(\lambda^3-1) - \lambda^{3}} \frac{\nu^{(\beta)} \breve{\mathcal G}^2 \overline{\gamma} \epsilon}{\mathcal{G}^{(1)} \mathcal{G}^{(2)}} ,
\end{equation}
where we have introduced the parameter $\epsilon = \delta^{(2)}-\delta^{(1)}$, which is assumed small. The first term, $\breve{\mathcal G} \overline{\gamma} / \mathcal{G}^{(\alpha)}$, is the expression of the shear strains in special Mooney--Rivlin materials ($\epsilon = 0$) or under simple shear ($\lambda=1$). Next, we substitute these expressions into Eq.~\eqref{InvarValid}, and we form the linear average of the energies \eqref{WMiniDeBotton}. Thus, we find
\begin{equation}
	\begin{aligned}
	\widetilde{W} &\simeq \tfrac12 \langle\mathcal{G}(1-\delta)\rangle (\lambda^2 + 2\lambda^{-1} - 3) % \\
	% &\quad
	+ \tfrac12 \langle\mathcal{G}\delta\rangle (\lambda^{-2} + 2\lambda - 3) \\
	&\quad + \tfrac12 \breve{\mathcal{G}}^2 \langle \mathcal{G}^{-1} (1-\delta) \rangle \overline{\gamma}^2 + \tfrac12 \breve{\mathcal{G}}^2 \langle \mathcal{G}^{-1} \delta \rangle \overline{\gamma}^2 \lambda^{-1} ,
	\end{aligned}
\end{equation}
up to terms of order $\epsilon^2$ or higher.
From these calculations as well as comparison with Eqs.~\eqref{WMREff} and \eqref{EffValid}, we deduce that the effective theory is exact in simple shear ($\lambda=1$). Furthermore, it is at least asymptotically correct for small $\epsilon$, as expected by construction. Nevertheless, we note that for the present deformations \eqref{FValid}, the effective theory described by Eq.~\eqref{WMREff} holds for any value of $\delta^{(1)} \simeq \delta^{(2)}$, without assuming that these parameters are small. Hence, the validity of this effective theory can go beyond the assumption $\delta^{(1)} \simeq \delta^{(2)} \ll 1$ that was made in Section~\ref{sec:WeffMRPert}.

%%%%%%%%%%%%%%%%%%%%%%%%%%%%%%%%%%%%%%%%%%%%%%%%%%%%%%%
\subsection{Pure shear}
%%%%%%%%%%%%%%%%%%%%%%%%%%%%%%%%%%%%%%%%%%%%%%%%%%%%%%%

Let us consider deformations that are restricted to a plane normal to a given unit vector, say ${\bf e}_3$. Thus, the macroscopic deformation gradient tensor $\overline{\bf F}$ has only four independent components in the orthonormal coordinate system defined by the basis vectors $\lbrace {\bf e}_1, {\bf e}_2, {\bf e}_3\rbrace$ (to be specified). In general, according to the polar decomposition theorem, the deformation gradient may be represented with respect to the Lagrangian principal axes by the product between a rotation tensor and a diagonal stretch tensor \citep[Chap. 2.2.6]{ogden84}. Here, one may introduce the stretch ratio $\lambda > 0$ along the first principal axis of deformation and the angle $\varphi$ between the Lagrangian and Eulerian principal axes. With respect to the chosen coordinate system, we find
\begin{equation}
	\begin{aligned}
	\overline{\bf F} &= \begin{bmatrix}
		\cos(\theta-\varphi) & -\sin(\theta-\varphi) & 0 \\
		\sin(\theta-\varphi) & \cos(\theta-\varphi) & 0\\
		0 & 0& 1
	\end{bmatrix} \begin{bmatrix}
		\lambda & 0 & 0 \\
		0 & \lambda^{-1} & 0\\
		0 & 0& 1
	\end{bmatrix} \dots \\
	&\quad\dots \begin{bmatrix}
		\cos\theta & \sin\theta & 0 \\
		-\sin\theta & \cos\theta & 0\\
		0 & 0& 1
	\end{bmatrix} , \end{aligned} 
	\label{FStretch}
\end{equation}
where we have introduced the rotation matrix of angle $\theta$ that transforms $\lbrace {\bf e}_1, {\bf e}_2\rbrace$ into the principal axes of deformation, see also \citet[Chap. 2.6]{holzapfel00} for useful identities.

The macroscopic deformation represented by the tensor \eqref{FStretch} with $\lambda = 1$ corresponds to a rigid-body rotation of angle $-\varphi$ about the origin, whereas the deformation gradient tensor \eqref{FStretch} with $\varphi = 0$ corresponds to a deformation known as \emph{pure shear}, or strip-biaxial extension. Furthermore, the special angle $\theta = -\frac12 \arctan (2/\kappa)$ corresponds to a simple shear deformation of ratio $\kappa = \lambda - \lambda^{-1}$, see also \citet{debotton05}. In what follows, we consider pure shear deformations for which $\varphi=0$, i.e.,
\begin{equation}
	\overline{\bf F} = \begin{bmatrix}
		\lambda \cos^2\!\theta + \lambda^{-1} \sin^2\!\theta & (\lambda - \lambda^{-1}) \cos\theta\sin\theta & 0 \\
		(\lambda - \lambda^{-1}) \cos\theta\sin\theta & \lambda^{-1} \cos^2\!\theta + \lambda \sin^2\!\theta & 0\\
		0 & 0& 1
	\end{bmatrix} .
	\label{FPureShear}
\end{equation}
This deformation gradient tensor has unit determinant, consistently with the incompressibility assumption, and the invariants $\tilde{I}_1$, $\tilde{I}_2$ are both equal to $\lambda^2 + \lambda^{-2} + 1$.

In Figure~\ref{fig:Stretch}, we represent the effect of this pure shear deformation on a square domain for specific values of $\theta$ and $\lambda$. Here, the material is extended with a stretching ratio $\lambda=1.5$ along the direction of polar angle $\theta=\pi/4$. Simultaneously, the material is compressed with a stretching ratio of $\lambda^{-1} \approx 0.66$ along the direction of polar angle $\theta=-\pi/4$. If the lamination direction is along ${\bf N}$ in the undeformed medium, then the lamination direction in the deformed medium is given by the vector ${\bf m}$ introduced in Eq.~\eqref{VecN}. These observations are completely independent of the actual orientation of the layers or on their mechanical behaviour.

To validate our asymptotic homogenisation results, we set the lamination direction to ${\bf N} = {\bf e}_1$, so that
\begin{equation}
	\tilde I_4 = \lambda^2 \cos^2\!\theta + \lambda^{-2} \sin^2\!\theta , \quad
	\tilde I_5 = \lambda^{-2} \cos^2\!\theta + \lambda^{2} \sin^2\!\theta .
\end{equation}
Based on Eqs.~\eqref{IdDE}-\eqref{IdKL}, we observe that $E=D$ and $L=K$,
for the present family of deformations. A minimiser $\bm\theta$ of the energy \eqref{WMiniDeBotton} is necessarily of the form $[\theta_2 \overline{\rm F}_{12}/\overline{\rm F}_{22}, \theta_2, \theta_3]$ where the macroscopic deformation satisfies Eq.~\eqref{FPureShear}. This way, the orthogonality constraint ${\bm \theta} \cdot {\bf m} = 0$ is satisfied.

In what follows, we use Matlab's \texttt{fminsearch} function to minimise the energy \eqref{WMiniDeBotton} with respect to the pair $(\theta_2, \theta_3)$ of independent components of $\bm \theta$. The default parameters are used in the minimisation algorithm. The initial guess is set to the exact solution ${\bm\theta} = a {\bf u}$ deduced from the neo-Hookean model, see Eq.~\eqref{SolSimpl}.

\begin{figure*}
	\begin{minipage}{0.49\textwidth}
		\centering
		(a)
		
		\vspace{0.7em}
		
		\includegraphics{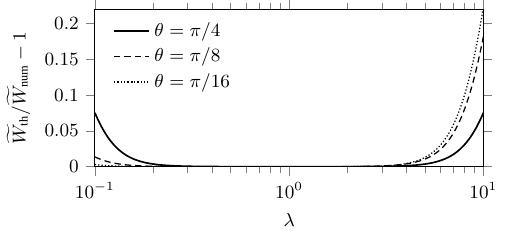}
	\end{minipage}\hfill
	\begin{minipage}{0.49\textwidth}
		\centering
		(b)
		
		\includegraphics{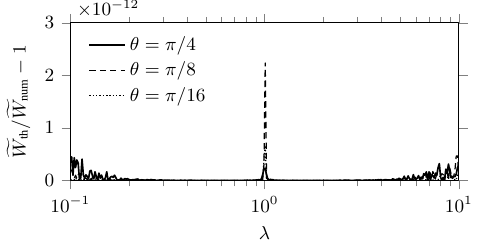}
	\end{minipage}
	\caption{ Validation of the asymptotic homogenisation results under pure shear \eqref{FPureShear}, for several values of the stretching ratio $\lambda$ and stretching angle $\theta$. (a) Yeoh laminates ($\nu^{(1)} = 0.8 = 1-\nu^{(2)}$, $\mathcal{G}^{(1)} = 4.3~\text{MPa} = \mathcal{G}^{(2)}/4$, $b^{(1)} = 0.013$, $b^{(2)} = 0.005$, $d^{(1)} = 2 b^{(1) 2}$, $d^{(2)} = b^{(2) 2}$). Relative error in the energy function $\widetilde{W}_\text{th}$ deduced from the analytical expression \eqref{WGenNH} with respect to a numerical minimisation $\widetilde{W}_\text{num}$ of the energy \eqref{WMiniDeBotton}. (b) Mooney--Rivlin laminates (same $\nu^{(\alpha)}$ and $\mathcal{G}^{(\alpha)}$ as above, $\delta^{(1)} = 0.8$, $\delta^{(2)} = 0.4$). Evolution of the relative error in the analytical expression \eqref{WMREff} of the effective energy. \label{fig:Valid}}
\end{figure*}

\paragraph{Yeoh laminates.} We first investigate the accuracy of Eq.~\eqref{WGenNH} for Yeoh laminates under pure shear. Here, we select the stretching angles $\theta = \pi/4$, $\pi/8$ and $\pi/16$, as well as 300 values of the stretching ratio $0.1 \leq \lambda \leq 10$ that are equally spaced on a logarithmic scale. Fig.~\ref{fig:Valid}a shows the evolution of the relative error between Eq.~\eqref{WGenNH} and a numerical solution to the constrained minimisation problem \eqref{WMiniDeBotton}. It can be observed that the estimated error does not exceed 5\% for the stretching ratios $0.2 < \lambda < 6$, which illustrates the accuracy of Eq.~\eqref{WGenNH} over a wide range of pure shear deformations. The chosen parameter values are of the same order of magnitude as those used to describe elastic deformations in rubber samples{\,---\,}typical parameter values are presented in more detail later on (see Eq.~\eqref{Param}). Here, the relative error shown in Fig.~\ref{fig:Valid}a could be reduced further by taking the corrections \eqref{WGenNHCorr} into account (not shown here). Nevertheless, the high accuracy already provided by the main terms \eqref{WGenNH} in the expansion of the energy may be deemed satisfactory in this specific application.

\paragraph{Mooney--Rivlin laminates.} Let us now investigate the accuracy of Eq.~\eqref{WMREff} for Mooney--Rivlin laminates under pure shear. Since $E=D$ and $L=K$, it appears that the homogenised energy \eqref{WMREff} derived asymptotically for Mooney--Rivlin laminates takes the same values as the exact one obtained for neo-Hookean laminates under pure shear. To estimate potential errors numerically, we cover the same range of stretching ratios as above. Fig.~\ref{fig:Valid}b shows the evolution of the relative error between Eq.~\eqref{WMREff} and a numerical solution to the constrained minimisation problem \eqref{WMiniDeBotton}. It can be observed that the estimated error is very small (of order $10^{-12}$) for the present deformations, which demonstrates the high accuracy of Eq.~\eqref{WMREff} under pure shear, even for values of the parameters $\delta^{(\alpha)}$ that are neither close nor small. Nevertheless, the accuracy of this homogenised energy function might be limited for other deformations.

%%%%%%%%%%%%%%%%%%%%%%%%%%%%%%%%%%%%%%%%%%%%%%%%%%%%%%%
\section{Incremental macro-deformations under pure shear}\label{sec:Insta}
%%%%%%%%%%%%%%%%%%%%%%%%%%%%%%%%%%%%%%%%%%%%%%%%%%%%%%%
\subsection{Macroscopic stability analysis}
%%%%%%%%%%%%%%%%%%%%%%%%%%%%%%%%%%%%%%%%%%%%%%%%%%%%%%%

To illustrate the main characteristics of the effective strain energy functions proposed in this study (Section~\ref{sec:Weff}), we consider incremental deformations of rank-1 layered composites under uniaxial planar deformations. In this context, the stability of neo-Hookean laminates was investigated by \citet{triantafyllidis85} based on a macroscopic and a microscopic stability analysis, and the equivalence between the emergence of instability at both scales was established. Further results on the macroscopic and microscopic stability of Gent laminates can be found in the study by \citet{yao24}.\textsuperscript{\ref{foot:yao}} The stability of sequentially layered neo-Hookean materials is discussed by \citet{debotton05}, whereas the stability of compressible neo-Hookean laminates is analysed by means of a second-order homogenisation procedure by \citet{lopezpamies09}. Using similar approaches, the stability of electro-active or magneto-active neo-Hookean laminates was studied by several authors, see \citet{spinelli15} and related works.

To obtain the relevant macroscopic stability conditions, we assume that the layered medium is subjected to a pre-deformation ${\bf X} \mapsto {\bf x}$ which is homogeneous and described by the effective deformation gradient tensor $\overline{\bf F}$. Furthermore, the mechanical response of the layered medium is described by the effective strain energy function $\widetilde{W}$. The incremental response of such an incompressible elastic solid is governed by the equations \citep{ogden84,ogden07}
\begin{equation}
	\mathbb{C}_{ijkl} \dot{\rm  x}_{k,jl} = \dot{p}_{,i} , \quad
	\dot{\rm x}_{i,i} = 0,
	\label{Incr}
\end{equation}
where indices after a comma indicate spatial differentiation with respect to the components of $\bf x$. Thus, $\dot{\rm x}_{k,l}$ and $\dot p_{,i}$ stand for $\partial \dot{\rm  x}_{k} / \partial  {\rm x}_l$ and $\partial \dot{p} / \partial {\rm x}_i$, respectively, where $\dot{\bf x}$ and $\dot p$ are the infinitesimal increments in the deformation and in the Lagrange multiplier used to enforce incompressibility. The fourth-order tensor of incremental elastic moduli is given by
\begin{equation}
	\mathbb{C}_{ijkl} = \overline{\rm F}_{ja} \frac{\partial^2 \widetilde{W} }{ \partial \overline{\rm F}_{ia} \partial \overline{\rm F}_{kb} } \overline{\rm F}_{lb} = \mathbb{C}_{klij} ,
	\label{StiffTens}
\end{equation}
whose components are calculated in the Appendix~\ref{app:Acoust}. The symmetry property of the tensor $\mathbb{C}$ above follows from the equality of mixed partials.
Here, we have used Einstein's summation convention for repeated indices. The first equation of \eqref{Incr} expresses the incremental balance of linear momentum, whereas the second one expresses the incremental constraint of incompressibility.

We seek $\dot{\bf x}$ in the form ${\bf q}\, f({\bf p}\cdot {\bf x})$ where $\bf p$ and $\bf q$ are unit vectors and $f$ is an arbitrary function. A similar assumption for the incremental pressure is made, $\dot{p} = g({\bf p}\cdot {\bf x})$. Upon substitution in the incremental equations \eqref{Incr}, we may rewrite the latter in the form
\begin{equation}
	{\bm \Gamma} {\bf q} f'' = g' {\bf p}, \quad 
	{\bf q} f' \cdot {\bf p}  = 0 ,
	\label{Acoust}
\end{equation}
where the primes $^\prime$ denote differentiation. Here, we have introduced the \emph{acoustic tensor} ${\bm \Gamma}$ whose components are given by $\Gamma_{ij} = \mathbb{C}_{iIjJ} {\rm p}_I {\rm p}_J$. The latter satisfies the symmetry property $\Gamma_{ij} = \Gamma_{ji}$, which is a consequence of the symmetry \eqref{StiffTens} of $\mathbb{C}$. If we define the projection tensor $\check{\bf I} = {\bf I} - {\bf p}\otimes {\bf p}$ onto the plane normal to $\bf p$, then we note that ${\bf q} = \check{\bf I} {\bf q}$ due to incompressibility. This way, Eq.~\eqref{Acoust} yields the restriction $\check{\bm \Gamma}\, {\bf q} = {\bf 0}$ where $\check{\bm \Gamma} = \check{\bf I} {\bm \Gamma} \check{\bf I}$ is a symmetric tensor.

Among potential solutions to the algebraic equation $\check{\bm \Gamma}\, {\bf q} = {\bf 0}$, the trivial solution ${\bf q} = {\bf 0}$ always exists. The existence of non-trivial elements in the kernel of $\check{\bm \Gamma}$ can be prevented by enforcing the \emph{strong ellipticity condition}
\begin{equation}
	{\bf q} \cdot \check{\bm \Gamma}\, {\bf q} > 0 ,
	\label{EllipticityCond}
\end{equation}
for all ${\bf q}$ orthogonal to an arbitrary direction $\bf p$. Under this condition, the projection $\check{\bm \Gamma}$ of the acoustic tensor is positive definite on the plane orthogonal to $\bf p$. Conversely, macroscopic instabilities may develop as soon as the strong ellipticity condition is no longer satisfied.

In what follows, we investigate the macroscopic stability of layered media subjected to pure shear deformations within a plane normal to ${\bf e}_3$.
Since we are restricting ourselves to plane isochoric deformations within this plane, it appears that the increments $\dot{\bf x}$ have only two independent components in the present coordinate system, namely $\dot{\rm x}_1$ and $\dot{\rm x}_2$. In addition, for a very large body, we note that the incremental problem \eqref{Incr} is macroscopically invariant along ${\bf e}_3$. This way, the unit vectors ${\bf p}$ and $\bf q$ introduced in Eq.~\eqref{Acoust} belong to the ${\bf e}_1$-${\bf e}_2$ plane.

%%%%%%%%%%%%%%%%%%%%%%%%%%%%%%%%%%%%%%%%%%%%%%%%%%%%%%%
\subsection{Macroscopic stability results}
%%%%%%%%%%%%%%%%%%%%%%%%%%%%%%%%%%%%%%%%%%%%%%%%%%%%%%%
\subsubsection{Preliminaries}
%%%%%%%%%%%%%%%%%%%%%%%%%%%%%%%%%%%%%%%%%%%%%%%%%%%%%%%

In this study, we focus on soft laminates of Yeoh type \eqref{WYeoh3} (and by extension, other generalised neo-Hookean materials \eqref{WGent}), or of Mooney--Rivlin type \eqref{WMRPhase}. In order to produce physically meaningful results, we consider a combined energy function of the form
\begin{equation}
	\begin{aligned}
	W^{(\alpha)} &= -\frac{(1-\delta) \mathcal{G}^{(\alpha)}}{2 b} \ln \left(1 - b (I_1^{(\alpha)}-3)\right) %\\
	%& \quad 
	+ \frac{\delta \mathcal{G}^{(\alpha)}}2 (I_2^{(\alpha)}-3) ,
	\end{aligned}
	\label{WCombined}
\end{equation}
where $\mathcal{G}^{(\alpha)}$ is the shear modulus of the phase $\alpha$, and the dimensionless parameters $\delta^{(\alpha)}=\delta$ and $b^{(\alpha)}=b$ are the same in both phases. This way, the Gent model corresponds to $\delta=0$, the (special) Mooney--Rivlin model \eqref{WMRSimpl} is recovered for $b \to 0$, while neo-Hookean laminates are obtained for $\delta=0$ and $b \to 0$.
For illustrative purposes, the physical parameters of the phase $\alpha=1$ are chosen according to the analysis of the Treloar data by \citet[Table~4]{destrade17} based on the same model, which leads us to setting
\begin{equation}
	\mathcal{G}^{(1)} = 4.3~\text{MPa} = \mathcal{G}^{(2)}/\rho, \quad 
	\delta=0.5, \quad 
	b=0.013,
	\label{Param}
\end{equation}
unless specified otherwise. For later use, we have introduced the shear modulus contrast $\rho = \mathcal{G}^{(2)}/\mathcal{G}^{(1)}$ which is a positive parameter.

\begin{figure}
	\centering
	\begin{minipage}{0.49\textwidth}
		\centering
		
		\includegraphics{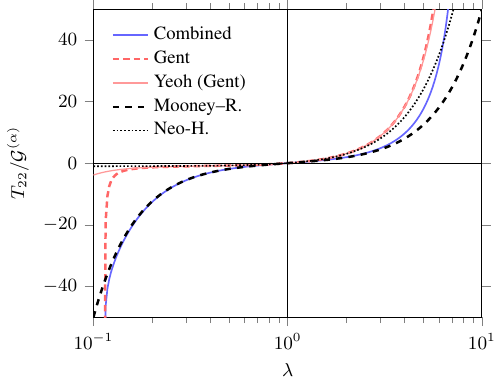}
	\end{minipage}
	\caption{ Homogeneous medium \eqref{WCombined} with the parameters \eqref{Param} and $\rho=1$. Stress-stretch relationship \eqref{StressStretch} for a pure shear deformation of stretch ratio $\lambda$ applied along the direction ${\bf e}_2$. The material is subjected to the stretch $\lambda^{-1}$ along the direction ${\bf e}_1$, and the other boundaries are assumed free. The Gent material corresponds to $\delta=0$ whereas the Mooney--Rivlin material is obtained for $b\to 0$. The Yeoh (Gent) material corresponds to Eq.~\eqref{WYeoh3} with $b^{(\alpha)}=b$ and $k^{(\alpha)}=2$. \label{fig:Stress}}
\end{figure}

The stress-stretch relationship for such a material is shown in Figure~\ref{fig:Stress}, where a homogeneous and isotropic material is considered ($\rho = 1$). The material is subjected to a pure shear deformation \eqref{FPureShear} for which stretching occurs in the direction of angle $\theta=\pi/2$, without loss of generality. The diagonal components $T_{ii}$ of the Cauchy stress tensor $\bf T$ are deduced from Eq.~\eqref{StressMicro} with the strain energy function \eqref{WCombined}, and we have
\begin{equation}
	\frac{T_{22}}{\mathcal{G}^{(\alpha)}} = \frac{(1-\delta) (\lambda^{2} -1)}{1-b (\lambda^2 + \lambda^{-2} - 2)} - \delta (\lambda^{-2} -1) .
	\label{StressStretch}
\end{equation}
The stress component $T_{11}$ is deduced from the above formula by replacing $\lambda$ with $\lambda^{-1}$.
Here, we have assumed vanishing tractions $T_{33}$ on the boundaries with unit normal ${\bf e}_3$ to eliminate the Lagrange multiplier. In the right-hand side of Fig.~\ref{fig:Stress}, significant differences at large stretches $\lambda > 1$ can be observed when the material model \eqref{WCombined} with the parameters \eqref{Param} is compared to the Gent model ($\delta=0$) and to the Mooney--Rivlin model ($b\to 0$). However, some of these discrepancies are significantly reduced in compression ($\lambda < 1$), or equivalently, in the dependence of $T_{11}$ with respect to $\lambda$ at large stretches (left-hand side of Fig.~\ref{fig:Stress}).

In general, layered media described by the strain energy function \eqref{WCombined} do not belong to the special class of laminates considered in Section~\ref{sec:WeffSimpl} since the ratio $R^{(\alpha)} = W^{(\alpha)}_2 / W^{(\alpha)}_1$ is not a constant (here, $R^{(\alpha)}$ is a function of $I_1^{(\alpha)}$). Nevertheless, based on earlier results, we can postulate an approximate effective energy function of the form
\begin{equation}
	\widetilde{W} = (1-\delta) \widetilde{W}_\text{YG} + \delta \left( \tfrac12 \langle \mathcal{G} \rangle E + \tfrac12 \breve{\mathcal G}  L \right) ,
	\label{WCombinedEff}
\end{equation}
which is assumed valid for small $b$. Here, the energy contribution $\widetilde{W}_\text{YG}$ that corresponds to the limit $\delta = 0$ is obtained from the expression of the effective energy \eqref{WGenNH} for Gent laminates, which are approximated as three-term Yeoh laminates \eqref{WYeoh3} with the parameters $b^{(\alpha)} = b$ and $d^{(\alpha)} = 2 b^2$ in both phases{\,---\,}see Fig.~\ref{fig:Stress}, where the evolution of the stress deduced from these models is displayed. Here, we neglect the corrections \eqref{WGenNHCorr}. The effective energy function for special Mooney--Rivlin laminates \eqref{WMRSimplEff} is recovered for $b\to 0$.

%%%%%%%%%%%%%%%%%%%%%%%%%%%%%%%%%%%%%%%%%%%%%%%%%%%%%%%
\subsubsection{Validation}
%%%%%%%%%%%%%%%%%%%%%%%%%%%%%%%%%%%%%%%%%%%%%%%%%%%%%%%

\begin{figure}
	\centering
	
	\includegraphics{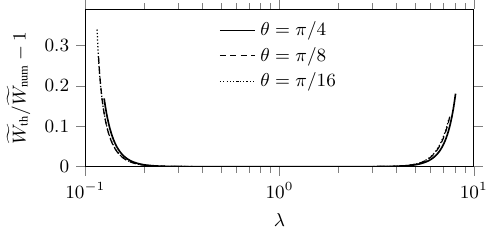}
	
	\caption{ Validation of the postulated energy function \eqref{WCombinedEff} under pure shear \eqref{FPureShear}, with the parameter values \eqref{Param}, $\nu^{(1)} = 0.8$, and $\rho=4$. Relative error in the energy function $\widetilde{W}_\text{th}$ deduced from the analytical expression \eqref{WCombinedEff} with respect to a numerical minimisation $\widetilde{W}_\text{num}$ of the energy \eqref{WMiniDeBotton}. \label{fig:ValidComb}}
\end{figure}

Let us investigate the accuracy of Eq.~\eqref{WCombinedEff} for laminates described by the energy function \eqref{WCombined} with the parameters \eqref{Param}. As shown in Fig.~\ref{fig:Stress}, the Gent energy function \eqref{WGent}\textsubscript{1} is well approximated by the three-term Yeoh energy function \eqref{WYeoh3} with $d^{(\alpha)}=2 b^{(\alpha) 2}$ for stretches ranging from 0.2 to 6. However, it remains to determine the accuracy of the postulated energy function \eqref{WCombinedEff} by comparing it to reference solutions deduced from the constrained minimisation problem \eqref{WMiniDeBotton} with the phase-related energies \eqref{WCombined}. Here, these computations are carried out based on the same procedure as in Section~\ref{sec:Valid}, and the results are shown below.

We begin with homogeneous deformations such that ${\bf u} = {\bf 0}$, for which $K=L=0$, and $D= \tilde I_1-3$, $E = \tilde I_2-3$. Thus, upon calculating the average energy \eqref{WMiniDeBotton} deduced from Eq.~\eqref{WCombined} and expanding the result as a Taylor series in terms of the small parameter $b$, we obtain the same estimate of the energy as provided by Eq.~\eqref{WCombinedEff}. Hence, the postulated effective energy function is asymptotically correct for such deformations.

Next, we consider simple shear motions described by the deformation gradient tensors \eqref{FValid} where $\lambda = 1$. We solve the system formed by Eq.~\eqref{RelShear} and by the definition of the macroscopic shear strain $\overline{\gamma}$ asymptotically, for small values of the parameter of nonlinearity, $b$. At leading order in $b$, we find that the shear strains in each phase satisfy Eq.~\eqref{ShearYeohValid} with $b^{(\alpha)} = b^{(\beta)} = b$. Substitution in Eqs.~\eqref{InvarValid}, \eqref{WCombined} and \eqref{WMiniDeBotton} yields
\begin{equation}
	\widetilde{W} \simeq (1-\delta) \left( \tfrac12 \breve{\mathcal G} \overline{\gamma}^2 + \tfrac14 \breve{\mathcal G}^4 \langle \mathcal{G}^{-3} b \rangle \overline{\gamma}^4 \right) + \delta \left(\tfrac12 \breve{\mathcal G} \overline{\gamma}^2 \right) ,
\end{equation}
up to an error of order $b^2$ or higher. Comparison with Eq.~\eqref{WCombinedEff} together with Eq.~\eqref{EffValid} shows that the postulated energy function \eqref{WCombinedEff} is coherent with this result, at least at the present order of approximation. Furthermore, the errors are dominated by those arising from the effective Gent/Yeoh behaviour (through $\widetilde{W}_\text{YG}$), since the Mooney--Rivlin part of Eq.~\eqref{WCombinedEff} with $b\to 0$ is exact in simple shear (cf. Section~\ref{sec:Valid}).

Figure~\ref{fig:ValidComb} represents the evolution of the error introduced by the estimate \eqref{WCombinedEff} of the macroscopic energy under pure shear \eqref{FPureShear}. Here, the reference $\widetilde{W}_\text{num}$ is obtained by numerical resolution of the constrained minimisation problem \eqref{WMiniDeBotton} with the energies \eqref{WCombined}. Upon comparison with the errors in Fig.~\ref{fig:Valid} obtained for Yeoh laminates and for Mooney--Rivlin laminates, we can observe that the error resulting from the postulated effective energy \eqref{WCombinedEff} is dominated by the Gent/Yeoh error contribution (the Mooney--Rivlin error contribution is negligible under pure shear). Furthermore, by construction, the effective energy \eqref{WCombinedEff} does not account for strain-limiting behaviour, i.e., for the restriction $I_1^{(\alpha)}-3 < b^{-1}$ due to the logarithm in Eq.~\eqref{WCombined}. This feature is shown in Fig.~\ref{fig:ValidComb} through the limited range of stretches accessible by the numerical solution with the layer response \eqref{WCombined}. Nevertheless, as shown in the figure, the relative error introduced by Eq.~\eqref{WCombinedEff} does not exceed 5\% for $0.2 < \lambda < 6$ with the present parameter values.

%%%%%%%%%%%%%%%%%%%%%%%%%%%%%%%%%%%%%%%%%%%%%%%%%%%%%%%
\subsubsection{Macroscopic stability of the undeformed medium}
%%%%%%%%%%%%%%%%%%%%%%%%%%%%%%%%%%%%%%%%%%%%%%%%%%%%%%%

We derive conditions for the macroscopic stability of the undeformed medium in the Appendix~\ref{app:Acoust}, which corresponds to setting $\overline{\bf F} = {\bf I}$ in the strong ellipticity condition \eqref{EllipticityCond}. As stated in Eq.~\eqref{CondUndef}, for this property to hold, we need that both $2(\widetilde{W}_D + \widetilde{W}_E)$ and $2 (\widetilde{W}_K + \widetilde{W}_L )$ are positive once evaluated in the undeformed state. For the soft laminates described by the effective energy function \eqref{WCombinedEff}, we find
\begin{equation}
	2(\widetilde{W}_D + \widetilde{W}_E)|_{\overline{\bf F}={\bf I}} = \langle \mathcal{G} \rangle , \quad
	2(\widetilde{W}_K + \widetilde{W}_L)|_{\overline{\bf F}={\bf I}} = \breve{\mathcal G} ,
	\label{StabUndef}
\end{equation}
since $D$ and $K$ are equal to zero in the undeformed state. The same result as above is obtained for other types of layered media discussed in this study, including Mooney--Rivlin laminates \eqref{WMREff} in the first-order approximation.

Since the shear modulus $\mathcal{G}^{(\alpha)}$ in each phase is positive, we know that the arithmetic mean $\langle \mathcal{G} \rangle$ and the harmonic mean $\breve{\mathcal G}$ are both positive. We conclude that for all the effective strain energy functions presented in this study, the undeformed state is stable. This result is independent of the orientation of the layers. Furthermore, it holds independently of the parameter values describing the mechanical response of each phase.

%%%%%%%%%%%%%%%%%%%%%%%%%%%%%%%%%%%%%%%%%%%%%%%%%%%%%%%
\subsubsection{Onset of macroscopic instability}
%%%%%%%%%%%%%%%%%%%%%%%%%%%%%%%%%%%%%%%%%%%%%%%%%%%%%%%

Let us consider a state of pure shear \eqref{FPureShear}. In a similar fashion to \citet{lopezpamies09}, we assume that the lamination direction ${\bf N} = {\bf e}_1$ belongs to the plane in which the deformation is applied.\footnote{The case of out-of-plane layers such that ${\bf N} = {\bf e}_3$ was also investigated by the authors but no macro-instability was detected, consistently with known results obtained for neo-Hookean laminates (see Appendix~\ref{app:Acoust}).} We now aim to determine the value of a potential critical stretch $\lambda = \lambda_\text{c} > 1$ for which there exists a stretching angle $0 \leq \theta \leq \pi/2$ such that strong ellipticity \eqref{EllipticityCond} no longer holds, for a given pair of orthogonal unit vectors $({\bf p}, {\bf q})$ within the plane normal to ${\bf e}_3$. In the special case of neo-Hookean laminates for which $\delta=b= 0$, \citet{triantafyllidis85} found that macroscopic instability develops at the stretching angle $\theta = 0$ and
\begin{equation}
	\lambda_\text{c} = \left(1 - \frac{\breve{\mathcal{G}}}{\langle{\mathcal{G}}\rangle}\right)^{-\frac14} , \quad
	\frac{\breve{\mathcal{G}}}{\langle{\mathcal{G}}\rangle} = \frac{(\nu^{(1)} + (1-\nu^{(1)})/\rho)^{-1}}{\nu^{(1)} + (1-\nu^{(1)})\rho} ,
	\label{StabNH}
\end{equation}	
see the calculations in the Appendix~\ref{app:Acoust} where this result is recovered. Hence, for neo-Hookean laminates, the critical stretch depends on two independent parameters only, namely the shear modulus ratio $\rho= \mathcal{G}^{(2)}/\mathcal{G}^{(1)}$ and the volume fraction $\nu^{(1)}=1-\nu^{(2)}$ of one material.

For the effective strain energy function \eqref{WCombinedEff}, we implement a straightforward algorithm that provides an estimation of the critical stretch $\lambda_\text{c}$ based on the components of the acoustic tensor $\bm \Gamma$, whose expression is given in the Appendix~\ref{app:Acoust}. We describe the incremental motion through ${\bf p} = [\cos \vartheta, \sin \vartheta , 0]$ and ${\bf q} = [-\sin \vartheta, \cos \vartheta , 0]$. Starting from $\lambda = 1$, the stretching ratio is increased by steps of $0.1$ until the strong ellipticity condition is no longer satisfied. Given $\lambda > 1$, we minimise the quantity $f = {\bf q} \cdot \check{\bm \Gamma}\, {\bf q}$ numerically with respect to the angles $\theta \in [-\pi/2, \pi/2]$ and $\vartheta \in [0, \pi]$. Here, this minimisation is performed using Matlab's \texttt{fmincon} function, with the initial guess $(\theta,\vartheta) \approx (0.05,1.55)$ and a tolerance of $10^{-6}$ for the function value. The minimum function value $\min f$ estimated numerically is then compared to zero. If a change of sign is detected, say $\min f < 0$ for $\lambda = \lambda_\text{M}$, then we use a root-finding method to identify the critical stretch $\lambda_\text{c}$ such that $-0.1 < \lambda_\text{c} - \lambda_\text{M} < 0$ and $\min f \approx 0$. Here, we have used Matlab's \texttt{fzero} function with a tolerance of $10^{-5}$ on $\min f$.

\begin{figure*}
	\centering
	\begin{minipage}{0.49\textwidth}
		\centering
		(a)
		
		\vspace{0.3em}		
				
		\includegraphics{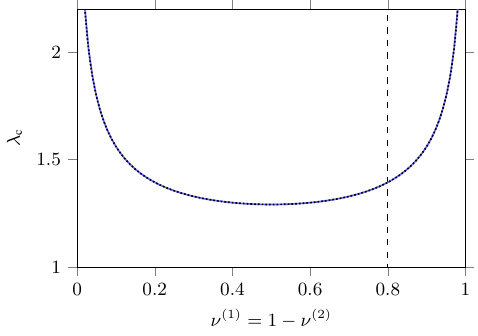}
	\end{minipage}\hfill	
	\begin{minipage}{0.49\textwidth}
		\centering
		(b)
		
		\includegraphics{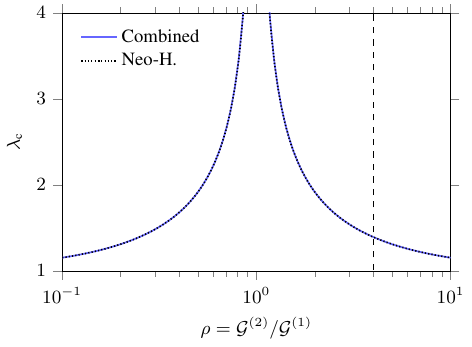}
	\end{minipage}
	\caption{Evolution of the critical stretch (a) in terms of the shear modulus ratio $\rho$ for $\nu^{(1)}=0.8$ (a logarithmic scale is used for $\rho$), and (b) in terms of the volume fraction $\nu^{(1)}$ for $\rho=4$. The vertical dashed lines mark the values $\rho=4$ and $\nu^{(1)}=0.8$, respectively. The curves were obtained for the combined model \eqref{WCombinedEff} with the parameters \eqref{Param}, and in the neo-Hookean limit \eqref{StabNH} ($\delta = b= 0$). \label{fig:Stab}}
\end{figure*}

Using this algorithm, we can investigate the influence of the shear modulus contrast $\rho$ and of the aspect ratio $\nu_1$ on the critical stretch for several values of $b$ and $\delta$. Consistently with known results obtained for neo-Hookean laminates (see Eq.~\eqref{StabNH}), we can keep the reference shear modulus $\mathcal{G}^{(1)} = 4.3$~MPa unchanged since $\lambda_\text{c}$ does not depend on it explicitly in this case. In Figure~\ref{fig:Stab}a, we display the evolution of the critical stretch with respect to $\rho$ for $\nu_1 = 0.8$ using the parameters of Eq.~\eqref{Param}. Similarly, Fig.~\ref{fig:Stab}b shows the evolution of the critical stretch with respect to $\nu_1$ for $\rho = 4$ using the same set of parameters as in Fig.~\ref{fig:Stab}a. In both cases, numerical results show that macroscopic instability develops when the material is stretched along the direction $\theta \approx 0$ and for $\vartheta \approx \pi/2$, with very small numerical variations of the order of $10^{-8}$ in both angles. Furthermore, the magnitude of the strong ellipticity indicator $f$ obtained numerically does not exceed $10^{-8}$ at these critical angles and at the critical stretch. For $\nu_1 = 0.8$ and $\rho=4$, Eq.~\eqref{StabNH} yields $\lambda_\text{c} \approx 1.4$ and ${\breve{\mathcal{G}}}/{\langle{\mathcal{G}}\rangle} \approx 0.74$. It is worth noting that the stretching ratios shown in Fig.~\ref{fig:Stab} belong to the domain of validity of the homogenised theory \eqref{WCombinedEff}, as shown in Fig.~\ref{fig:ValidComb}.

In Figure~\ref{fig:Stab}, we observe that macroscopic instability develops at the same critical stretch in laminates whose layers are described by the combined energy function \eqref{WCombined}, or by a neo-Hookean energy function. Therefore, the dependence of the strain energy function on limiting chain extensibility or on the second strain invariant is visibly not very significant as far as the macroscopic stability of such soft layered media under pure shear is concerned. Nevertheless, for the present family of constitutive models, we have assumed that the parameters $b^{(\alpha)}$, $d^{(\alpha)}$, $\delta^{(\alpha)}$ are the same in both phases (they are equal to $b$, $2 b^2$, and $\delta$, respectively). Modifications of the macroscopic instability threshold might occur when the rescaled energy functions $W^{(1)}/\mathcal{G}^{(1)}$ and $W^{(2)}/\mathcal{G}^{(2)}$ differ.

To explore this possibility, one could examine layered media involving significant differences in these quantities (e.g., $b^{(1)}\neq b^{(2)}$, $d^{(1)}\neq d^{(2)}$, or $\delta^{(1)}\neq \delta^{(2)}$). For example, this could be achieved with a layered material that combines neo-Hookean layers and Yeoh layers with two terms in the polynomial expansion of the energy. The energy function $W^{(1)}$ of the neo-Hookean layers can be expressed by Eq.~\eqref{WYeoh3} with $b^{(1)}=d^{(1)}=0$, whereas the energy function $W^{(2)}$ of the Yeoh layers can be obtained from Eq.~\eqref{WYeoh3} with $d^{(2)}=0$ and $b^{(2)}=b\neq 0$. In this case, it appears that the rescaled energy functions $W^{(1)}/\mathcal{G}^{(1)}$ and $W^{(2)}/\mathcal{G}^{(2)}$ deduced from Eq.~\eqref{WYeoh3} have distinct expressions, contrary to the layered medium described by Eq.~\eqref{WCombined} in both phases. In the Appendix~\ref{app:Acoust}, preliminary results obtained for specific orientations suggest that the expression \eqref{StabNH} of the critical stretch might no longer hold in this case, see Eq.~\eqref{StabNE3NProp}.

%%%%%%%%%%%%%%%%%%%%%%%%%%%%%%%%%%%%%%%%%%%%%%%%%%%%%%%
\subsection{Wave propagation analysis}\label{subsec:Wave}
%%%%%%%%%%%%%%%%%%%%%%%%%%%%%%%%%%%%%%%%%%%%%%%%%%%%%%%

Incremental wave propagation can be analysed based on a `small-on-large' formalism that is similar to the one describing material stability. Such a framework was exploited in the study by \citet{galich17}, who showed that the wave propagation characteristics in soft laminates are highly tunable by pre-deformation. For time-dependent incremental motions, Eq.~\eqref{Incr} becomes
\begin{equation}
	\mathbb{C}_{ijkl} \dot{\rm  x}_{k,jl} - \dot{p}_{,i} = \varrho (\dot{\rm  x}_{i})_{tt} , \quad
	\dot{\rm x}_{i,i} = 0,
	\label{IncrWav}
\end{equation}
where the subscript $t$ indicates partial differentiation with respect to time, and $\varrho$ is the average mass density of the layered medium. Then, we seek $\dot{\bf x}$ in the form ${\bf q}\, f({\bf p}\cdot {\bf x} - ct)$ where $({\bf p}, {\bf q})$ are unit vectors, $c$ represents the wave speed, and a similar assumption is made for the incremental pressure, $\dot{p} = g({\bf p}\cdot {\bf x}-ct)$. Upon substitution in Eq.~\eqref{IncrWav}, we find
\begin{equation}
	f'' {\bm \Gamma} {\bf q} - g' {\bf p} = \varrho c^2 {\bf q}, \quad 
	f' {\bf q} \cdot {\bf p}  = 0 ,
	\label{AcoustWav}
\end{equation}
where ${\bm \Gamma}$ is the acoustic tensor.
Eq.~\eqref{AcoustWav} yields the eigenvalue problem $\check{\bm \Gamma} {\bf q} = \varrho c^2 {\bf q}$ with $\check{\bm \Gamma} = \check{\bf I} {\bm \Gamma} \check{\bf I}$, see Eq.~\eqref{Acoust} for a definition of these quantities. If $\check{\bm \Gamma}$ is positive definite (i.e., if the strong ellipticity condition \eqref{EllipticityCond} is satisfied), then the eigenvalues $\varrho c^2$ are positive, and the wave speeds $c$ are real.

To analyse the effect of the pre-deformation on the effective wave propagation characteristics, we assume that the material is subjected to in-plane extension along the layer interfaces. Thus, if the lamination direction is ${\bf N} = {\bf e}_1$, then we set $\theta = \pi/2$ in Eq.~\eqref{FPureShear}. With this choice, we expect that macroscopic instability might develop in compression only ($\lambda < 1$), in agreement with the analysis of stability provided in the previous section. Next, we consider $401$ angles $\vartheta$ that are equally spaced over $[-\pi,\pi]$ to describe the incremental motion through the propagation vector ${\bf p} = [\cos \vartheta, \sin \vartheta , 0]$ and the polarisation vector ${\bf q} = [-\sin \vartheta, \cos \vartheta , 0]$. Given a value of the stretching ratio $\lambda$, we can evaluate the quantity ${\bf q} \cdot \check{\bm \Gamma}\, {\bf q} = \varrho c^2$ based on the expression of the acoustic tensor (Appendix~\ref{app:Acoust}). This procedure allows us to produce the slowness curves, i.e., the polar plots of $c^{-1} \sqrt{\langle \mathcal{G} \rangle / \varrho}$ in terms of the angle $\vartheta$.

\subsubsection{Results for proportional laminates}

Before discussing the wave propagation properties further, let us reassess the accuracy of our approximate effective theory \eqref{WCombinedEff} with proportional energies, i.e., the parameters $\delta^{(\alpha)} = \delta$, $b^{(\alpha)} = b$, and $d^{(\alpha)} = 2 b^2$ are the same in both phases. We consider a specific propagation angle, $\vartheta=0$, which corresponds to normal incidence with respect to the layer interfaces (${\bf p} = {\bf N} = {\bf e}_1$). The bi-axial tensile motion described by Eq.~\eqref{FPureShear} with $\theta = \pi/2$ is homogeneous throughout the layered medium. In the effective medium, the quantity ${\bf q}\cdot \check{\bm\Gamma} {\bf q}$ is given by Eq.~\eqref{StabNE10} with $\lambda$ replaced by $\lambda^{-1}$, i.e.,
\begin{equation}
	\varrho c^2 = \breve{\mathcal G} \lambda^{-2} \left( (1-\delta) \left(1 + b D + b^2 D^2\right) + \delta \right) ,
	\label{SpeedCombEff}
\end{equation}
where $D = \lambda^2 + \lambda^{-2} - 2$.

Within a layer described by the strain energy function \eqref{WCombined}, the components of the acoustic tensor provided in the Appendix~\ref{app:Acoust} yield the expression of the wave speed $c^{(\alpha)}$ under the same state of deformation,
\begin{equation}
	\varrho^{(\alpha)} (c^{(\alpha)})^2 = \mathcal{G}^{(\alpha)} \lambda^{-2} \left( \frac{1-\delta}{1-b(\lambda^2 + \lambda^{-2} - 2)} + \delta \right) ,
	\label{SpeedComb}
\end{equation}
where $\varrho^{(\alpha)}$ is the mass density in the phase $\alpha$. Based on this expression, we can estimate the effective wave speed in the layered medium at normal incidence \citep{berjamin25},
\begin{equation}
	\varrho c^2 = \breve{\mathcal G} \lambda^{-2} \left( \frac{1-\delta}{1-b(\lambda^2 + \lambda^{-2} - 2)} + \delta \right) ,
	\label{SpeedCombHom}
\end{equation}
where $\varrho = \nu^{(1)}\varrho^{(1)} + \nu^{(2)}\varrho^{(2)}$ is the average mass density of the layered medium. By comparison with Eq.~\eqref{SpeedCombEff}, we observe that the error introduced by the approximation \eqref{WCombinedEff} of the macroscopic energy amounts to the error in the polynomial approximation of the logarithm in Eq.~\eqref{WCombined}, at cubic order. This conclusion is coherent with other validations performed for homogeneous deformations in Section~\ref{sec:Valid}.

\begin{figure*}
	\centering
	\begin{minipage}{0.49\textwidth}
		\centering
		
		(a)
		
		\includegraphics{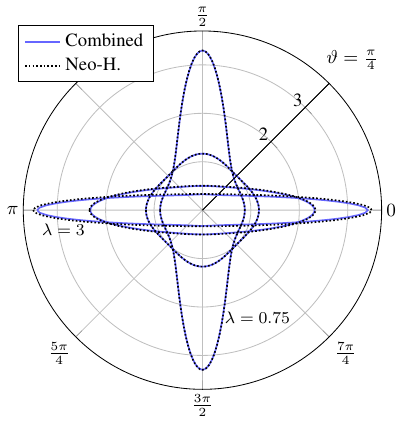}
	\end{minipage}\hfill
	\begin{minipage}{0.49\textwidth}
		\centering
		
		(b)
		
		\includegraphics{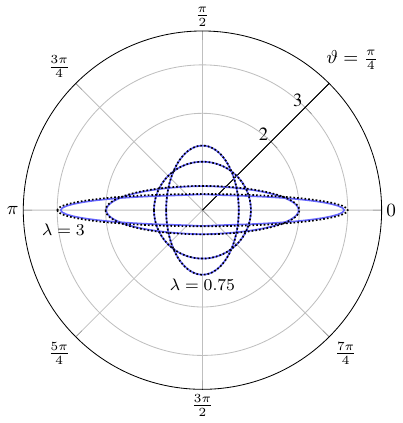}
	\end{minipage}
	\caption{ (a) Slowness curves for shear waves propagating in a soft laminate of stiffness ratio $\rho = 4$ and volume fraction $\nu^{(1)} = 0.8$. The waves are polarised within the ${\bf e}_1$-${\bf e}_2$ plane, and they propagate in this plane at an angle $\vartheta$ with respect to the lamination direction, ${\bf N} = {\bf e}_1$. The quantity displayed in polar coordinates is $c^{-1} \sqrt{\langle \mathcal{G} \rangle / \varrho}$ where $c$ is the wave speed deduced from the effective model \eqref{WCombinedEff} with the parameters \eqref{Param}. The material is subjected to in-plane stretching \eqref{FPureShear} along the layer interfaces ($\theta=\pi/2$), with stretching ratios $\lambda = 0.75, 1, 2, 3$. (b) Same graphics obtained for a homogeneous medium where $\rho = 1$ and $\nu^{(1)} = 1$. \label{fig:Polar}}
\end{figure*}

Let us now produce the slowness curves, which are shown in Figure~\ref{fig:Polar}a. Here, we have selected soft laminates described by the energy function \eqref{WCombinedEff}, with stiffness ratio $\rho = 4$ and volume fraction $\nu^{(1)} = 0.8$. The effective mass density is $\varrho = 10^3$~kg/m\textsuperscript{3}, and the remaining material parameters are taken from Eq.~\eqref{Param}. It should be noted that these parameter values were obtained from quasi-static experiments; they may therefore not be very accurate for analysing the wave propagation properties in real media. In any case, the effective response of the layered medium can be deduced from Eq.~\eqref{WCombinedEff}. At moderate stretching ratios ($\lambda \simeq 1$), there is no visible difference in terms of wave speed between the neo-Hookean layered medium and the layered medium that combines Mooney--Rivlin elasticity with Yeoh elasticity \eqref{WCombinedEff}.

It can be observed that the wave speed in the Mooney--Rivlin--Yeoh medium is slightly larger than in the neo-Hookean medium at large stretches $\lambda=3$ for the angle $\vartheta = 0$.
This observation is coherent with the expression of the effective wave speed at normal incidence. In fact, the relative difference between the squared wave speed \eqref{SpeedCombEff} and its neo-Hookean counterpart is of order $(1-\delta)  b D$, which corresponds to approximately 5\% for the present parameter values \eqref{Param} and $\lambda=3$. The neo-Hookean wave speed is deduced from Eq.~\eqref{SpeedCombEff} with $\delta=0$ and $b=0$.

The same slowness curves obtained for a homogeneous medium are shown in Fig.~\ref{fig:Polar}b. Here, we have used the same procedure as for Fig.~\ref{fig:Polar}a but with the stiffness ratio $\rho = 1$ and the volume fraction $\nu^{(1)} = 1$. Based on these results, it can be observed that the pre-deformation has a comparable effect on wave speeds in the layered medium with proportional phases and in the homogeneous medium. In other words, for the present constitutive models and parameter values, the effect of the constitutive choices \eqref{WCombined} on the slowness curves is moderate, and this effect can be observed at large stretches only.

\begin{figure}
\centering
	\begin{minipage}{0.49\textwidth}
		\centering
		
		\includegraphics{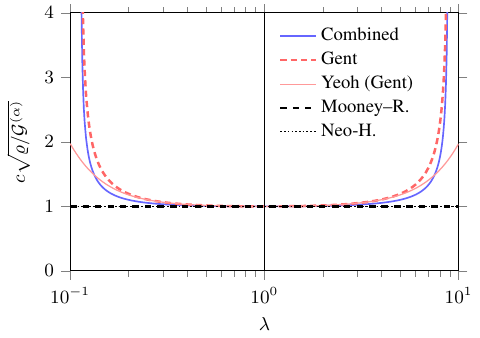}
	\end{minipage}
	
\caption{ Rescaled wave speed in a homogeneous medium under pure shear \eqref{FPureShear} at the angle $\theta=\pi/2$. The waves are polarised in the stretching direction and they propagate orthogonally to it ($\vartheta = 0$). The same models and parameters as in Fig.~\ref{fig:Stress} are used. \label{fig:SpeedHom} }
\end{figure}

This feature is illustrated again in Figure~\ref{fig:SpeedHom} where we compare values of the wave speed obtained for several homogeneous media. For a homogeneous medium described by the strain energy function \eqref{WCombined}, the wave speed satisfies Eq.~\eqref{SpeedComb} under pure shear, with $\theta=\pi/2$ and $\vartheta=0$.
The same calculations are performed for all the models considered in Fig.~\ref{fig:Stress}, which include limiting cases of Eq.~\eqref{SpeedComb}, as well as the three-term Yeoh model \eqref{WYeoh3}. The parameter values are set according to Eq.~\eqref{Param} and $d^{(\alpha)}=2 b^{(\alpha) 2}$.

The evolution of the wave speed with respect to the stretching ratio for various constitutive models is shown in Fig.~\ref{fig:SpeedHom}, at selected angles. In contrast with the significant differences between models that can be observed in Fig.~\ref{fig:Stress} (which displays the evolution of the stress components with respect to the stretching ratio), only subtle differences between models can be observed in Fig.~\ref{fig:SpeedHom}. This radical difference between the evolution of the stress components and of the wave speed in terms of the pre-deformation can be seen directly by comparing Eqs.~\eqref{StressStretch} and \eqref{SpeedComb}. Nevertheless, even though it may not seem obvious at first glance, it is possible to establish a relationship between pre-stress components and wave speeds. In fact, as shown through Eq.~\eqref{Gower}, the difference between two pre-stress components is linked to the speed of two waves propagating in perpendicular directions.

\subsubsection{Results for non-proportional laminates}

Let us briefly investigate wave propagation in a non-proportional layered medium that combines neo-Hookean layers with two-term Yeoh layers. The strain energy in the layers is described by Eq.~\eqref{WYeoh3} with $b^{(1)}=0$, $b^{(2)}=b$, and $d^{(1)}=d^{(2)}=0$. The material's effective response is deduced from Eq.~\eqref{WYeoh}, or equivalently, Eq.~\eqref{WGenNH}. Thus, upon substituting $\lambda$ with $\lambda^{-1}$ in Eq.~\eqref{StabNE105NProp}, the relationship \eqref{SpeedCombEff} satisfied by the effective wave speed becomes
\begin{equation}
	\varrho c^2 = \breve{\mathcal G} \lambda^{-2} \left(1  + \nu^{(2)} b D \breve{\mathcal G}/ {\mathcal G}^{(2)}  \right)  ,
	\label{SpeedNPropEff}
\end{equation}
under pure bi-axial shear at normal incidence ($\theta=\pi/2$, $\vartheta=0$). Once the wave speeds \eqref{SpeedComb} in each phase have been calculated, the wave speed at normal incidence can be obtained in a similar fashion as Eq.~\eqref{SpeedCombHom}, i.e.,
\begin{equation}
	\varrho c^2 = \breve{\mathcal G}\lambda^{-2} \left( 1 - \frac{\nu^{(2)}\breve{\mathcal G}}{\mathcal{G}^{(2)}} \frac{b D}{1+bD} \right)^{-1} .
\end{equation}
These expressions are asymptotically equivalent for small values of the parameter of nonlinearity $b$, which confirms the accuracy of the effective energy \eqref{WYeoh3} at the present order of approximation.

In the present case, the relative difference between the squared wave speed \eqref{SpeedNPropEff} and its neo-Hookean counterpart is of order $\nu^{(2)} b D \breve{\mathcal G}/ {\mathcal G}^{(2)}$. For the parameter values \eqref{Param} and $\lambda=3$, we have $b D \approx 10\%$ and $\nu^{(2)} \breve{\mathcal G}/ {\mathcal G}^{(2)} \approx 6 \%$, where we have set $\nu^{(1)} = 0.8$ and $\rho = 4$ (i.e., the neo-Hookean phase is in excess, and it is the softer one). Thus, for this specific choice, it turns out that the wave speed is very close to the value obtained in the corresponding neo-Hookean laminate. However, this outcome could be clearly modified by assuming that the Yeoh phase is in excess, and that this phase is the softer one. For instance, if we set $\nu^{(1)} = 0.2$ and $\rho = 1/4$, then $\nu^{(2)} \breve{\mathcal G}/ {\mathcal G}^{(2)} \approx 94 \%$. This way, the relative difference between the squared wave speed \eqref{SpeedNPropEff} and its neo-Hookean counterpart would reach about 9\%, a difference that is much more significant than in the proportional case \eqref{SpeedCombEff}.

%%%%%%%%%%%%%%%%%%%%%%%%%%%%%%%%%%%%%%%%%%%%%%%%%%%%%%%
\section{Concluding remarks}\label{sec:Conclu}
%%%%%%%%%%%%%%%%%%%%%%%%%%%%%%%%%%%%%%%%%%%%%%%%%%%%%%%

Using homogenisation theory, we have provided analytical expressions of the effective strain energy in periodic layered media made of rubber-like material. At the macroscopic scale, the material described by this type of effective strain energy function is transversely isotropic. Eqs.~\eqref{WMRSimplEff}, \eqref{WGenNH} and \eqref{WMREff} generalise a known result obtained for neo-Hookean layers to a more general class of constitutive laws (see \citet{debotton05} for the neo-Hookean case), which includes Yeoh layers and Mooney--Rivlin layers (Table~\ref{tab:Expr}). These results can be used to model soft composite materials more accurately than with the neo-Hookean model, as shown in the literature through the comparison between these models and experimental measurements carried out on rubber samples in uniaxial tension \citep{destrade17}, or for other deformations \citep{horgan25}.

While the homogenisation result \eqref{WMRSimplEff} for special Mooney--Rivlin layers is exact, this is not the case for the homogenisation results \eqref{WGenNH} and \eqref{WMREff} obtained asymptotically for Yeoh layers and for Mooney--Rivlin layers. The accuracy of these expressions is confirmed after validation under various deformations, at least asymptotically. The accuracy of the postulated energy function \eqref{WCombinedEff} describing layered media with Mooney--Rivlin and Gent-type behaviour was assessed in the same manner. The results reported in this study were obtained within a range of high accuracy of the proposed homogenised models.

We have used our results to highlight the effect of these constitutive assumptions on the macroscopic instability threshold of a periodic soft laminate subjected to pure shear. Using the Gent hyperelastic model (in the form of a three-term Yeoh model) and the Mooney--Rivlin model, we found that the effect of these constitutive choices on the critical stretch is negligible in soft composites when a certain proportionality relationship is satisfied by the elastic energy of the constituents (Fig.~\ref{fig:Stab}). A similar observation is made for the speed of incremental shear waves that propagate in a pre-deformed laminate under pure shear (Fig.~\ref{fig:Polar}). In fact, we found that the effect of limiting chain extensibility and of Mooney--Rivlin behaviour are of the same order of magnitude in homogeneous media as in stratified media for which this proportionality relationship holds.

Both of these observations seem to be closely linked to the proportionality assumption $W^{(1)}/\mathcal{G}^{(1)} = W^{(2)}/\mathcal{G}^{(2)}$. Preliminary results obtained in non-proportional cases show that the constitutive response of individual layers could have a significant impact on the macroscopic instability threshold, or on the speed of shear waves propagating at normal incidence, see Eqs.~\eqref{SpeedNPropEff} and \eqref{StabNE105NProp}-\eqref{StabNE3NProp}. More general properties may be associated with the above proportionality relationship, but such implications remain to be explored in future works.

At this stage, it is worth recalling the main limitations of the present study, in which the layers are assumed incompressible, isotropic, and purely elastic. In addition, the composite medium is assumed periodic, with perfectly bonded layers, and the usual assumptions underpinning homogenisation theory are made. Furthermore, most of the results presented in this study were obtained for a layered medium in which the strain energies \eqref{WCombined} of the constituents satisfy the relationship $W^{(1)}/\mathcal{G}^{(1)} = W^{(2)}/\mathcal{G}^{(2)}$, with the exception of some preliminary results mentioned above. Moreover, it is equally important to note that the parameter values \eqref{Param} were obtained from tensile experimental data \citep{destrade17}, and not from wave propagation tests. Consequently, it is possible that other stratified media would lead to quantitatively different predictions, within the same homogenisation framework.

The results obtained here could be exploited to improve the quality of theoretical predictions deduced from homogenised theories, or to adapt existing homogenisation frameworks to account for a larger class of constitutive laws \citep{lopezpamies09}. In particular, they could be used to analyse the macroscopic stability of magneto-active and electro-active composites \citep{spinelli15}, the microscopic instability of layered media \citep{yao24},\textsuperscript{\ref{foot:yao}} mechanical test results \citep{horgan25}, or the wave propagation characteristics in layered media at normal incidence \citep{berjamin25}. Furthermore, the proposed effective energy functions could be used as a building block to develop generalised continuum models that describe the dynamic response of composite media, e.g., within the Cosserat theory or the micromorphic theory \citep{munch11,forest20}. Such theories could be elaborated to describe wave propagation along arbitrary directions (i.e., at oblique incidence), see for instance \citet{delph78} for the dispersion of small-amplitude waves, as well as \citet{ketcheson20} for homogenised nonlinear wave equations. For potential generalisations to elastoplastic, viscoelastic and magneto-active materials, the interested reader is referred to the works by \citet{agoras26} and by \citet{ruggieri25}.

%%%%%%%%%%%%%%%%%%%%%%%%%%%%%%%%%%%%%%%%%%%%%%%%%%%%%%%
\subsection*{CRediT authorship contribution statement}
%%%%%%%%%%%%%%%%%%%%%%%%%%%%%%%%%%%%%%%%%%%%%%%%%%%%%%%

{\bfseries Harold Berjamin}: Writing -- original draft, Visualization, Methodology, Investigation, Formal analysis, Conceptualization. {\bfseries Stephan Rudykh}: Writing -- review \& editing, Supervision, Methodology, Funding acquisition, Conceptualization.

%%%%%%%%%%%%%%%%%%%%%%%%%%%%%%%%%%%%%%%%%%%%%%%%%%%%%%%
\subsection*{Declaration of competing interest}
%%%%%%%%%%%%%%%%%%%%%%%%%%%%%%%%%%%%%%%%%%%%%%%%%%%%%%%

The authors declare that they have no known competing financial interests or personal relationships that could have appeared to influence the work reported in this paper.

%%%%%%%%%%%%%%%%%%%%%%%%%%%%%%%%%%%%%%%%%%%%%%%%%%%%%%%
\subsection*{Declaration of generative AI use}
%%%%%%%%%%%%%%%%%%%%%%%%%%%%%%%%%%%%%%%%%%%%%%%%%%%%%%%

The authors report generative AI was not used in their research or preparation of this manuscript.

%%%%%%%%%%%%%%%%%%%%%%%%%%%%%%%%%%%%%%%%%%%%%%%%%%%%%%%
\subsection*{Acknowledgements}
%%%%%%%%%%%%%%%%%%%%%%%%%%%%%%%%%%%%%%%%%%%%%%%%%%%%%%%

This work was supported by the European Research Council (ERC) under Grant No. 852281 -- MAGIC.
HB is grateful to Doireann O'Kiely (University of Limerick) for fruitful discussions and support.
The authors thank the anonymous reviewers for their insightful and constructive comments.

\section*{Appendix}
\appendix

%%%%%%%%%%%%%%%%%%%%%%%%%%%%%%%%%%%%%%%%%%%%%%%%%%%%%%%
\section{Components of the acoustic tensor}\label{app:Acoust}
%%%%%%%%%%%%%%%%%%%%%%%%%%%%%%%%%%%%%%%%%%%%%%%%%%%%%%%

%%%%%%%%%%%%%%%%%%%%%%%%%%%%%%%%%%%%%%%%%%%%%%%%%%%%%%%
\subsection{General expression}
%%%%%%%%%%%%%%%%%%%%%%%%%%%%%%%%%%%%%%%%%%%%%%%%%%%%%%%

Let us calculate the components of the projected acoustic tensor $\check{\bm \Gamma}$, which is deduced from the acoustic tensor $\Gamma_{ij} = \mathbb{C}_{iIjJ} {\rm p}_I {\rm p}_J$ by left- and right-multiplication through the projection tensor $\check{\bf I} = {\bf I} - {\bf p}\otimes {\bf p}$, and where the coefficients $\mathbb{C}_{ijkl}$ are defined in Eq.~\eqref{StiffTens}. For $\widetilde{W}$ expressed in terms of the invariants $\tilde I_\alpha$ of Eq.~\eqref{InvarMacro}, we can write \citep{ogden07}
\begin{equation}
	\mathbb{C}
	=
	\widetilde{W}_\alpha \mathbb{D}_{\alpha}
	+ \widetilde{W}_{\alpha\beta}\, {\bm \Xi}_{\alpha} \otimes {\bm \Xi}_{\beta} , 
	\label{StiffDerGen}
\end{equation}
where
\begin{equation}
	\mathbb{D}_{\alpha|ijkl} = \overline{\rm F}_{ja} \frac{\partial^2 \tilde I_\alpha }{ \partial \overline{\rm F}_{ia} \partial \overline{\rm F}_{kb} } \overline{\rm F}_{lb}, \quad
	{\Xi}_{\alpha|ij} = \frac{\partial \tilde I_\alpha }{ \partial \overline{\rm F}_{ia} } \overline{\rm F}_{ja} ,
\end{equation}
are the components of the tensors $\mathbb{D}_\alpha$ and ${\bm \Xi}_\alpha$, and we have used Einstein's summation convention for repeated indices ($\alpha,\beta = 1,2,4,5$). The coefficients $\widetilde{W}_\alpha = \partial \widetilde{W} / \partial \tilde I_\alpha$ and $\widetilde{W}_{\alpha\beta} = \partial^2 \widetilde{W} / \partial \tilde I_\alpha \partial \tilde I_\beta$ are partial derivatives of the strain energy function with respect to the invariants. According to the equality of mixed partials, we have $\widetilde{W}_{\alpha\beta} = \widetilde{W}_{\beta\alpha}$.

If $\widetilde{W}$ is expressed in terms of the scalar functions $D$, $E$, $K$, $L$ introduced in Eqs.~\eqref{IdDE}-\eqref{IdKL}, then according to differentiation rules,
\begin{equation}
	\widetilde{W}_\alpha = \widetilde{w}_p \frac{\partial \tilde J_p}{\partial \tilde I_\alpha}, \quad
	\widetilde{W}_{\alpha\beta} = \widetilde{w}_p \frac{\partial^2 \tilde J_p}{\partial \tilde I_\alpha \partial \tilde I_\beta} + \widetilde{w}_{p q} \frac{\partial \tilde J_p}{\partial \tilde I_\alpha} \frac{\partial \tilde J_q}{\partial \tilde I_\beta} ,
\end{equation}
where $\tilde J_1 = D$, $\tilde J_2 = E$, etc. Here, we have introduced the coefficients $\widetilde{w}_p = \partial \widetilde{w}/\partial \tilde J_p$ and $\widetilde{w}_{p q} = \partial^2 \widetilde{w}/\partial \tilde J_p \partial \tilde J_q$. This way, using the notation $\widetilde{w}_1 = \widetilde{W}_D$, etc., we arrive at the formulas in Eq.~\eqref{RelW} for the first derivatives, and similar expressions can be obtained for the second derivatives.

Now, let us return to the computation of the stiffness tensor $\mathbb C$, see Eq.~\eqref{StiffDerGen}. Using the definition of the invariants $\tilde I_\alpha$ in Eq.~\eqref{InvarMacro}, we can readily compute their first and second partial derivatives with respect to $\overline{\bf F}$. The first derivatives satisfy
\begin{equation}
	\begin{aligned}
	&{\bm \Xi}_1 = 2 \overline{\bf B} , \quad {\bm \Xi}_2 = -2 \overline{\bf B}^{-1} , \quad
	{\bm \Xi}_4 = 2 \tilde I_4 {\bf n}\otimes {\bf n} ,\quad {\bm \Xi}_5 = -2 \tilde I_5 {\bf m}\otimes {\bf m} ,	
	\end{aligned}
	\label{StiffDer1}
\end{equation}
which are symmetric tensors. In particular, these expressions are used to calculate the macroscopic Cauchy stress tensor $\overline{\bf T} = \overline{\bf P}\overline{\bf F}^\text{T}$ in Eq.~\eqref{StressMacro}, which is equal to the sum $\overline{\bf T} = \widetilde{W}_\alpha {\bm \Xi}_\alpha - \overline{p}\, {\bf I}$. For the second derivatives, we find
\begin{equation}
	\begin{aligned}
		&\mathbb{D}_{1|ijkl} = \delta_{ik} {\Xi}_{1|jl} , \quad
		\mathbb{D}_{2|ijkl} = -\delta_{il} {\Xi}_{2|jk} - \delta_{jl} {\Xi}_{2|ik} - \delta_{jk} {\Xi}_{2|il} , \\
		&\mathbb{D}_{4|ijkl} = \delta_{ik} {\Xi}_{4|jl} , \quad
		\mathbb{D}_{5|ijkl} = -\delta_{il} {\Xi}_{5|jk} - \delta_{jl} {\Xi}_{5|ik} - \delta_{jk} {\Xi}_{5|il} ,
	\end{aligned}
	\label{StiffDer2}
\end{equation}
where $\delta_{ij}$ is the Kronecker delta.
From there, we deduce that the fourth-order incremental stiffness tensor reads
\begin{equation}
	\mathbb{C}_{ijkl} = \delta_{ik} \overline{\rm T}'_{jl}  - \delta_{il} \overline{\rm T}''_{jk} - \delta_{jl} \overline{\rm T}''_{ik} - \delta_{jk} \overline{\rm T}''_{il} + \widetilde{W}_{\alpha\beta}\, {\Xi}_{\alpha|ij} {\Xi}_{\beta|kl} ,
	\label{StiffTensExpr}
\end{equation}
where we have introduced the partial stresses $\overline{\bf T}' = \widetilde{W}_1 {\bm \Xi}_1 + \widetilde{W}_4 {\bm \Xi}_4$ and $\overline{\bf T}'' = \widetilde{W}_2 {\bm \Xi}_2 + \widetilde{W}_5 {\bm \Xi}_5$ such that the macroscopic Cauchy stress tensor of Eq.~\eqref{StressMacro} can be rewritten as $\overline{\bf T} = \overline{\bf T}' + \overline{\bf T}'' - \overline{p} \, {\bf I}$. If the only nonzero coefficients are the partial derivatives $\widetilde{W}_1$, $\widetilde{W}_4$ and $\widetilde{W}_{44}$, then the above formula \eqref{StiffTensExpr} reduces to Eq.~(32) of \citet{berjamin25b}.

Based on this result, the components of the acoustic tensor can be determined, and we find
\begin{equation}
	\bm{\Gamma} = ({\bf p}\cdot  \overline{\bf T}' {\bf p}) {\bf I}  - {\bf p} \otimes \overline{\bf T}'' {\bf p} - \overline{\bf T}'' {\bf p} \otimes {\bf p} - \overline{\bf T}'' + \widetilde{W}_{\alpha\beta}\, {\bm\Xi}_{\alpha} {\bf p} \otimes {\bm\Xi}_{\beta} {\bf p}  .
	\label{AcoustExpr}
\end{equation}
Left- and right-multiplication by the projection tensor $\check{\bf I}$ onto the plane orthogonal to $\bf p$ lends
\begin{equation}
	\check{\bm\Gamma} = ({\bf p}\cdot  \overline{\bf T}' {\bf p}) \check{\bf I} - \check{\bf I}\overline{\bf T}''\check{\bf I} + \widetilde{W}_{\alpha\beta}\, \check{\bf I} {\bm\Xi}_{\alpha} {\bf p} \otimes \check{\bf I} {\bm\Xi}_{\beta} {\bf p} ,
	\label{AcoustProjExpr}
\end{equation}
according to the property $\check{\bf I}{\bf p} = {\bf 0}$.
We note in passing that the above tensor $\check{\bm\Gamma}$ is symmetric, therefore it is diagonalisable with real eigenvalues. This includes the trivial eigenvalue $0$ for which $\bf p$ is an eigenvector.
Now, if we left- and right-multiply $\check{\bm\Gamma}$ by a unit vector ${\bf q}$ which is orthogonal to $\bf p$, we find
\begin{equation}
	{\bf q} \cdot \check{\bm\Gamma} {\bf q} = {\bf p}\cdot  \overline{\bf T}' {\bf p} - {\bf q} \cdot\overline{\bf T}'' {\bf q} + \widetilde{W}_{\alpha\beta}\, ({\bf q} \cdot {\bm\Xi}_{\alpha} {\bf p})  ({\bf q} \cdot {\bm\Xi}_{\beta} {\bf p}) ,
	\label{StabQuant}
\end{equation}
which, in passing, is also equal to ${\bf q} \cdot {\bm\Gamma} {\bf q}$. Strong ellipticity \eqref{EllipticityCond} requires this quantity to be positive.

The expression of ${\bf q} \cdot \check{\bm\Gamma} {\bf q}$ can be rewritten in a different form. As done in \citet{berjamin24}, we write Eq.~\eqref{StiffTensExpr} as follows,
\begin{equation}
	\mathbb{C}_{ijkl} = \delta_{ik} \left(\overline{\rm T}_{jl} + \overline{p}\delta_{jl}\right) + \mathbb{M}_{ijkl} ,
\end{equation}
where the fourth-order tensor defined by
\begin{equation}
	\mathbb{M}_{ijkl} = \widetilde{W}_{\alpha\beta}\, {\Xi}_{\alpha|ij} {\Xi}_{\beta|kl} - \delta_{ik}\overline{\rm T}''_{jl} - \delta_{il} \overline{\rm T}''_{jk} - \delta_{jl} \overline{\rm T}''_{ik} - \delta_{jk} \overline{\rm T}''_{il} 
\end{equation}
satisfies the major symmetry and the minor symmetries ($\mathbb{M}_{ijkl} = \mathbb{M}_{jikl} = \mathbb{M}_{ijlk} = \mathbb{M}_{klij}$). Using this notation, the acoustic tensor \eqref{AcoustExpr} becomes $\bm{\Gamma} = ({\bf p}\cdot \overline{\bf T} {\bf p} + \overline{p}) {\bf I} + \bm{\Lambda}$, together with the definition $\Lambda_{ij} = \mathbb{M}_{iIjJ} {\rm p}_I {\rm p}_J$ for the components of $\bm{\Lambda}$. The projection $\check{\bm\Gamma}$ of the acoustic tensor can be calculated in a similar fashion as in Eq.~\eqref{AcoustProjExpr}. From there, we find
\begin{equation}
	{\bf q} \cdot \check{\bm\Gamma} {\bf q} = {\bf p}\cdot \overline{\bf T} {\bf p} + \overline{p} + {\bf q} \cdot {\bm\Lambda} {\bf q} .
	\label{Gower}
\end{equation}
This formula can be used to estimate a stress difference based on two measurements of ${\bf q} \cdot \check{\bm\Gamma} {\bf q}$ (so as to eliminate the Lagrange multiplier $\overline{p}$ above), e.g., by measuring wave speeds along two well-chosen propagation directions \citep{berjamin24}.

%%%%%%%%%%%%%%%%%%%%%%%%%%%%%%%%%%%%%%%%%%%%%%%%%%%%%%%
\subsection{Undeformed medium}
%%%%%%%%%%%%%%%%%%%%%%%%%%%%%%%%%%%%%%%%%%%%%%%%%%%%%%%

Let us now consider the case where the material is in an undeformed state (i.e., $\overline{\bf F} = {\bf I}$). Under this assumption, we have
\begin{equation}
	\begin{aligned}
	{\bf q} \cdot \check{\bm\Gamma} {\bf q} &= 2(\widetilde{W}_1 + \widetilde{W}_2) + 2\widetilde{W}_4 {\rm p}_{\rm N}^2 + 2\widetilde{W}_5 {\rm q}_{\rm N}^2 \\
		&\quad	+ 4 \left(\widetilde{W}_{44} - 2\widetilde{W}_{45} + \widetilde{W}_{55}\right) {\rm p}_{\rm N}^2 {\rm q}_{\rm N}^2 , \\
	&= 2(1-\Phi) (\widetilde{W}_D + \widetilde{W}_E) + 2 \Phi (\widetilde{W}_K + \widetilde{W}_L) ,
	\end{aligned}
	\label{StabQuantUndef}
\end{equation}
where we have introduced $\Phi = {\rm p}_{\rm N}^2 + {\rm q}_{\rm N}^2 - 4 {\rm p}_{\rm N}^2 {\rm q}_{\rm N}^2$, as well as the projections ${\rm p}_{\rm N} = {\bf p}\cdot{\bf N}$ and ${\rm q}_{\rm N} = {\bf q}\cdot{\bf N}$ which are within $[-1 , 1]$. The partial derivatives of $\widetilde{W}$ above are evaluated in the undeformed state and we have used the relationships \eqref{RelW}. According to the Cauchy--Schwarz inequality, these quantities satisfy
\begin{equation}
	 {\rm p}_{\rm N}^2{\rm q}_{\rm N}^2 = |\widehat{\bf I}\, {\bf p}\cdot \widehat{\bf I}\, {\bf q}|^2 \leq |\widehat{\bf I}\, {\bf p}|^2 |\widehat{\bf I}\, {\bf q}|^2
	 = (1 - {\rm p}_{\rm N}^2) (1 - {\rm q}_{\rm N}^2) ,
\end{equation}
where $\widehat{\bf I}\, {\bf p} = {\bf p} - {\rm p}_{\rm N} {\bf N}$ is the projection of $\bf p$ onto the plane normal to $\bf N$ (the projection tensor $\widehat{\bf I}$ is here defined according to $\widehat{\bf I} = {\bf I} - {\bf N}\otimes {\bf N}$), and a similar definition is used for $\widehat{\bf I}\, {\bf q}$. Based on this inequality, we conclude that $0 \leq {\rm p}_{\rm N}^2 + {\rm q}_{\rm N}^2 \leq 1$.

Over this domain, it turns out that the polynomial function $\Phi$ takes values between zero and one. Therefore, for the positivity of \eqref{StabQuantUndef} to hold for any pair of orthogonal unit vectors $\bf p$ and $\bf q$, we request that
\begin{equation}
	2(\widetilde{W}_D + \widetilde{W}_E) > 0 , \quad
	2 (\widetilde{W}_K + \widetilde{W}_L ) > 0 ,
	\label{CondUndef}
\end{equation}
where all these partial derivatives are evaluated in the undeformed state. For the present family of energy functions $\widetilde{W}$ that satisfy the relationships \eqref{RelW}, this macroscopic stability condition is equivalent to the requirement
\begin{equation}
	2(\widetilde{W}_1 + \widetilde{W}_2)  > 0 , \quad
	2 \widetilde{W}_4 = 2 \widetilde{W}_5 > -2(\widetilde{W}_1 + \widetilde{W}_2) ,
	\label{CondUndefI}
\end{equation}
where all the terms are evaluated in the undeformed state ($\overline{\bf F} = {\bf I}$).

%%%%%%%%%%%%%%%%%%%%%%%%%%%%%%%%%%%%%%%%%%%%%%%%%%%%%%%
\subsection{Neo-Hookean laminates}
%%%%%%%%%%%%%%%%%%%%%%%%%%%%%%%%%%%%%%%%%%%%%%%%%%%%%%%

For neo-Hookean laminates described by the energy function \eqref{WMRSimplEff} with $\delta=0$, the only relevant coefficients in Eq.~\eqref{StabQuant} are
\begin{equation}
	\widetilde{W}_1 = \tfrac12 \langle \mathcal{G} \rangle, \quad
	\widetilde{W}_4 = \tfrac12 (\breve{\mathcal G} - \langle \mathcal{G} \rangle)= \tilde I_5^{2} \widetilde{W}_5 = -\tfrac12 \tilde I_5^{3}\widetilde{W}_{55}  .
	\label{DerNH}
\end{equation}
In particular, the inequalities \eqref{CondUndefI} are trivially satisfied, meaning that the undeformed state is stable.
Based on this expression of the partial derivatives of $\widetilde{W}$, it follows that the projected acoustic tensor \eqref{AcoustProjExpr} reads \citep{galich17}
\begin{equation}
	\check{\bm\Gamma} = c_1 \check{\bf I} + c_2 \check{\bf I} {\bf m} \otimes \check{\bf I} {\bf m} ,
\end{equation}
where
\begin{equation}
	\begin{aligned}
	&c_1 = \langle \mathcal{G} \rangle ({\bf p}\cdot  \overline{\bf B} {\bf p})  + (\breve{\mathcal G} - \langle \mathcal{G} \rangle) \tilde I_4 ({\bf p}\cdot  {\bf n})^2 , \\
	&c_2 = (\breve{\mathcal G} - \langle \mathcal{G} \rangle) \tilde I_5^{-1} \left(1 - 4 ({\bf p}\cdot {\bf m})^2\right) .
	\end{aligned}
\end{equation}
Noting that $\operatorname{tr}\check{\bm\Gamma} = 2c_1 + c_2 |\check{\bf I} {\bf m}|^2$ and that $\det\check{\bm\Gamma} = c_1^2 + c_1c_2 |\check{\bf I} {\bf m}|^2$ in the two-dimensional space normal to $\bf p$, we can obtain two relevant eigenspaces in this plane corresponding to the eigenvalues $c_1$ and $c_1 + c_2 |\check{\bf I} {\bf m}|^2$, where the relationship $|\check{\bf I} {\bf m}|^2 = 1 - ({\bf p}\cdot {\bf m})^2$ holds. Thus, the tensor $\check{\bm\Gamma}$ remains positive definite in this plane provided that the above eigenvalues are positive. Furthermore, we have \citep{agoras09}
\begin{equation}
	{\bf q}\cdot \check{\bm\Gamma} {\bf q} = c_1 + c_2 ({\bf q} \cdot {\bf m})^2 ,
\end{equation}
which is required to remain positive for every unit vector $\bf q$ orthogonal to $\bf p$, according to the strong ellipticity condition \eqref{EllipticityCond}.

For instance, for a macroscopic pure shear deformation \eqref{FPureShear} with stretching ratio $\lambda$ in the direction indicated by $\theta=0$, we can make the following observations:
\begin{itemize}[leftmargin=*]
	\item for ${\bf N} = {\bf e}_1$, any pair of orthonormal unit vectors $({\bf p},{\bf q})$ within the ${\bf e}_1$-${\bf e}_2$ plane leads to
\begin{equation}	
	\begin{aligned}
	{\bf q}\cdot \check{\bm\Gamma} {\bf q} & = \breve{\mathcal G} \lambda^{2} \cos^2\!\vartheta
	+ 4 \langle \mathcal{G} \rangle \lambda_\text{c}^{-4} \lambda^2 \cos^2\!\vartheta \sin^2\!\vartheta \\
	&\quad \phantom{0} + \langle \mathcal{G} \rangle \lambda^{-2} \left( 1 - \lambda^4/\lambda_\text{c}^{4}\right) \sin^2\!\vartheta ,
	\end{aligned}
	\label{EllipNH}
\end{equation}
where $\vartheta$ is the angle between $\bf p$ and the ${\bf e}_1$-axis, and
\begin{equation}
	\lambda_\text{c} = (1-\breve{\mathcal{G}}/\langle{\mathcal{G}}\rangle)^{-1/4} > 1, \quad
	\breve{\mathcal G} / \langle \mathcal{G} \rangle \leq 1 . 
	\label{LambC}
\end{equation}
Noting that the first two terms in Eq.~\eqref{EllipNH} are increasing non-negative functions of $\lambda$ and that the third term changes sign, we conclude that the present deformation becomes macroscopically unstable at the critical stretch $\lambda=\lambda_\text{c}$ for incremental displacements along the lamination direction ($\vartheta = \pi/2$). This result is highlighted in Eq.~\eqref{StabNH}, and it is coherent with the one obtained by \citet{triantafyllidis85}. The case ${\bf N} = {\bf e}_2$ can be tackled in a similar fashion, as it simply corresponds to a rotation of the coordinate axes.
	\item for ${\bf N} = {\bf e}_3$, any pair of orthonormal unit vectors $({\bf p},{\bf q})$ within the ${\bf e}_1$-${\bf e}_2$ plane produces ${\bf p} \cdot {\bf n} = 0$ and ${\bf q} \cdot {\bf m} = 0$. These properties lead to 
\begin{equation}	
	{\bf q}\cdot \check{\bm\Gamma} {\bf q} = \langle{\mathcal{G}}\rangle (\lambda^{2} \cos^2\!\vartheta + \lambda^{-2}\sin^2\!\vartheta) ,
	\label{NE3}
\end{equation}
where $\vartheta$ is the angle between $\bf p$ and the ${\bf e}_1$-axis. Since ${\bf q}\cdot \check{\bm\Gamma} {\bf q}$ is always positive, the deformation remains stable for incremental displacements transverse to the lamination direction (i.e., along the layer interfaces).
\end{itemize}
For completeness, this exercise could be reproduced for other stretching angles, $\theta \neq 0$. Strong ellipticity conditions obtained for other material models with transverse isotropy can be found in the works by \citet{merodio02,merodio05}.

%%%%%%%%%%%%%%%%%%%%%%%%%%%%%%%%%%%%%%%%%%%%%%%%%%%%%%%
\subsection{Special orientations}
%%%%%%%%%%%%%%%%%%%%%%%%%%%%%%%%%%%%%%%%%%%%%%%%%%%%%%%

We now provide some detailed analytical expressions obtained in special cases corresponding to certain stretching and polarisation directions. Here we consider the more general family of constitutive laws \eqref{WCombinedEff}, where the coefficients $b^{(\alpha)}=b$, $d^{(\alpha)}=2 b^2$, and $\delta^{(\alpha)}=\delta$ are not layer-dependent. The material is subjected to pure shear \eqref{FPureShear} of stretch $\lambda$ in the direction indicated by $\theta=0$. In this case, we find that
\begin{itemize}[leftmargin=*]
	\item for ${\bf N} = {\bf e}_1$, calculation of the quantity \eqref{StabQuant} can be cumbersome, in general. We will therefore restrict the result to the special cases $\vartheta = 0$ and $\vartheta = \pi/2$. In the first situation (${\bf p} = {\bf e}_1$, ${\bf q} = {\bf e}_2$), we find
\begin{align}	
	&{\bf q}\cdot \check{\bm\Gamma} {\bf q} = 2 (\widetilde{W}_1 + \widetilde{W}_2 + \widetilde{W}_4 ) \lambda^2 = 2 (\widetilde{W}_K + \widetilde{W}_L) \lambda^2, \label{StabNE10} \\
	&2 (\widetilde{W}_K + \widetilde{W}_L ) = (1-\delta) \breve{\mathcal G} \left(1 + b D + b^2 D^2\right) + \delta \breve{\mathcal G} ,
\end{align}
see Eqs.~\eqref{IdDE}-\eqref{IdKL} and \eqref{RelW}, where $D = \lambda^2 + \lambda^{-2} - 2$. In the second situation (${\bf p} = {\bf e}_2$, ${\bf q} = -{\bf e}_1$), we have
\begin{align}	
	& 	{\begin{aligned}
		{\bf q}\cdot \check{\bm\Gamma} {\bf q} &= 2 (\widetilde{W}_1 + \widetilde{W}_2 + \widetilde{W}_5 ) \lambda^{-2} %\\
			%&
			= 2 (\widetilde{W}_D + \widetilde{W}_E) \lambda^{-2} \left(1 - \lambda^4/\lambda_\text{c}^4\right) , \label{StabNE105} 
		\end{aligned}} \\
	&2 (\widetilde{W}_D + \widetilde{W}_E) = (1-\delta) \langle \mathcal{G} \rangle \left(1 + b D + b^2 D^2\right) + \delta \langle \mathcal{G} \rangle .
\end{align}
We note that the expression of ${\bf q}\cdot \check{\bm\Gamma} {\bf q}$ above changes sign at the critical stretch $\lambda = \lambda_\text{c}$ given by Eq.~\eqref{LambC}. This result is a consequence of the proportionality relationship $(\widetilde{W}_D + \widetilde{W}_E)/\langle{\mathcal G}\rangle = (\widetilde{W}_K + \widetilde{W}_L)/\breve{\mathcal G}$, which allows us to write the expression of ${\bf q}\cdot \check{\bm\Gamma} {\bf q}$ at $\vartheta = \pi/2$ in this specific form.
	\item for ${\bf N} = {\bf e}_3$, Eq.~\eqref{StabQuant} becomes
\begin{equation}	
	{\bf q}\cdot \check{\bm\Gamma} {\bf q} = 2 \widetilde{W}_1 ({\bf p}\cdot \overline{\bf B}{\bf p}) + 2 \widetilde{W}_2 ({\bf q}\cdot \overline{\bf B}^{-1}{\bf q}) + 4\widetilde{W}_{11} ({\bf q}\cdot \overline{\bf B}{\bf p})^2 ,
	\label{StabNE3}
\end{equation}
in general. Since each bracket yields a non-negative contribution, the above quantity remains non-negative provided that the coefficients $\widetilde{W}_1$, $\widetilde{W}_2$, $\widetilde{W}_{11}$ with $\widetilde{W} = \widetilde{W}_\text{iso}$ are non-negative, see Eq.~\eqref{WDec}. Based on Eqs.~\eqref{WDecMR} and \eqref{WDecGenNH}, it can be shown that this property is always true for the present constitutive law.
\end{itemize}

We now repeat these calculations for a material that combines a neo-Hookean layer and a two-term Yeoh layer, which is subjected to pure shear \eqref{FPureShear} of stretch $\lambda$ in the direction indicated by $\theta=0$. The energy functions $W^{(\alpha)}$ deduced from Eq.~\eqref{WYeoh3} have the following parameter values, $b^{(1)}=0$, $b^{(2)}=b$, and $d^{(1)}=d^{(2)}=0$. The material's effective response is deduced from Eq.~\eqref{WGenNH}. With this choice, we observe that the proportionality relationship $(\widetilde{W}_D + \widetilde{W}_E)/\langle{\mathcal G}\rangle = (\widetilde{W}_K + \widetilde{W}_L)/\breve{\mathcal G}$ does not hold as long as $b \neq 0$ and $\mathcal{G}^{(1)} \neq \mathcal{G}^{(2)}$, in general. Furthermore, we arrive at the following results:
\begin{itemize}[leftmargin=*]
	\item for ${\bf N} = {\bf e}_1$ and $\vartheta = 0$, the first identity in Eq.~\eqref{StabNE10} still holds. For the present material model, we find
\begin{equation}
	{\bf q}\cdot \check{\bm\Gamma} {\bf q} = 2 \widetilde{W}_K \lambda^2, \quad 
	2 \widetilde{W}_K = \breve{\mathcal G} + \nu^{(2)} \breve{\mathcal G}^2 (\mathcal{G}^{(2)})^{-1} b D , \label{StabNE105NProp}
\end{equation}
where $D = \lambda^2 + \lambda^{-2} - 2$, which is always non-negative.
For ${\bf N} = {\bf e}_1$ and $\vartheta = \pi/2$, the first identity in Eq.~\eqref{StabNE105} still holds, and we have
\begin{align}	
	& {\bf q}\cdot \check{\bm\Gamma} {\bf q} = 2\widetilde{W}_D \lambda^{-2} \left(1 - \lambda^4/\lambda_\text{c}^4 \right) , \\
	& 2\widetilde{W}_D = \langle{\mathcal G}\rangle + \nu^{(2)} {\mathcal G}^{(2)} b D , \quad \lambda_\text{c} = (1-\widetilde{W}_K/\widetilde{W}_D)^{-1/4} . \label{StabNE3NProp}
\end{align}
For $b = 0$, these expressions are consistent with the neo-Hookean result \eqref{StabNH} where $\vartheta = 0$ or $\vartheta = \pi/2$, respectively. We note that the expression of ${\bf q}\cdot \check{\bm\Gamma} {\bf q}$ above changes sign at the critical stretch $\lambda=\lambda_\text{c}$ whose value is determined by solving the nonlinear equation $1-\lambda^{-4} = \widetilde{W}_K/\widetilde{W}_D$, where $\widetilde{W}_K/\widetilde{W}_D$ depends on $D = \lambda^2 + \lambda^{-2} - 2$, in general. Thus, we expect a modification of the macroscopic instability threshold with respect to the neo-Hookean case \eqref{LambC} for the present class of stratified composites.
	\item for ${\bf N} = {\bf e}_3$, Eq.~\eqref{StabNE3} still holds. The same conclusion as for the previous material can be drawn.
\end{itemize}

%%%%%%%%%%%%%%%%%%%%%%%%%%%%%%%%%%%%%%%%%%%%%%%%%%%%%%%
{\footnotesize
\bibliography{Biblio}
}
%%%%%%%%%%%%%%%%%%%%%%%%%%%%%%%%%%%%%%%%%%%%%%%%%%%%%%%

\end{document}